\documentclass{aa}

\usepackage{graphicx}

\usepackage{natbib}
\usepackage{txfonts}

\begin{document}

\title {The chemical evolution of Barium and Europium in the Milky Way}

\author {G. Cescutti\inst{1}
\thanks {email to: cescutti@ts.astro.it}
\and  P. Fran\c cois\inst{2}
\and  F. Matteucci\inst{1}, \inst{4}
\and  R. Cayrel\inst{2}
\and  M. Spite\inst{3}}

\institute{ Dipartimento di Astronomia, Universit\'a di Trieste, via G.B. Tiepolo 11, I-34131  
\and  Observatoire de Paris/Meudon, GEPI, 61 Avenue de l'Observatoire, 75014 Paris, France
\and  Observatoire de Paris-Meudon, GEPI, 92195 Meudon Cedex, France
\and  I.N.A.F. Osservatorio Astronomico di Trieste, via G.B. Tiepolo 11, I-34131}

\date{Received xxxx / Accepted xxxx}

\abstract{}
{We compute the evolution of the abundances of barium and europium
in the Milky Way  and we compare our  results with
the observed abundances from the recent  UVES Large Program
"First Stars".}
{We use a chemical evolution model which already reproduces the majority of 
observational constraints.}
{We confirm that barium is a neutron capture 
element mainly produced in the low mass AGB stars during the 
thermal-pulsing phase by
the $^{13}C$ neutron source, in a slow neutron capture process. 
However, in order to reproduce the [Ba/Fe] vs. [Fe/H] as well as the Ba 
solar abundance, we suggest that
Ba should be also produced as an r-process element by massive stars in 
the range 10-30$M_{\odot}$.
On the other hand, europium should be only an r-process element  
produced in the same range of masses 
(10-30$M_{\odot}$), at variance with previous suggestions indicating a 
smaller mass range for the Eu producers.
As it is well known, there is a large spread in the [Ba/Fe] and [Eu/Fe] 
ratios at low metallicities, although smaller in the newest data. 
With our model we estimate for both elements (Ba and Eu) the ranges for the
r-process yields from massive stars which  better reproduce
the trend of the data.
  We find that with the same yields which are able to explain the 
observed trends, the 
large spread in the [Ba/Fe] and [Eu/Fe] ratios  cannot  be 
explained  even in the context of an inhomogeneous models
 for the chemical evolution of our Galaxy. 
We therefore derive the amount by which the yields should be modified 
to fully account for the observed spread.
We then discuss several possibilities to explain the size of the  spread.
We finally suggest that the production ratio of [Ba/Eu] could be
 almost constant in the massive stars.}{}


\maketitle
\section{Introduction}
The neutron capture is the main mechanism to form elements heavier than iron. The other
mechanism, the p-process, is required for the proton rich isotopes, whose abundances in the
solar system is less than 1\%. 
Two major neutron capture mechanisms are generally invoked: the slow process (s-process) and the rapid process
(r-process), where the slow and rapid are defined relatively to the timescale of $\beta$-decay.\\
The s-process requires a relatively low neutron density and moves along the valley of
$\beta$ stability. The s-process feeds in particular the elements Sr-Y-Zr, 
Ba-La-Ce-Pr-Nd and Pb, the three major abundance s-peaks.
The reason for the existence of these peaks is the following:
the neutron capture process imposes certain features on the "spectrum" 
of the heavy element abundances.
For certain neutron numbers N =  50, 82, 126 the neutron capture cross-sections 
are much
smaller than for neighbouring neutron numbers. This means that once one of 
these "magic" numbers
is reached, it becomes significantly less likely for the nucleus to capture 
more neutrons.
These numbers are a quantum mechanical effect of closed shells, in 
precisely the same way
that closed electron shells produce high chemical stability in the noble 
gases. 
Therefore, if the neutron capture process operates in some environment for 
some finite length of time and then shuts off, we expect a fair number of 
nuclei to be "stuck" at these "magic"
numbers. 
Elements which correspond to these "magic" numbers of neutrons will thus 
be especially abundant. We identify then three peaks, 
as described above.

The site of production of the s-elements is not unique. In fact, the
main component, accounting for the s-process in the atomic mass number 
range $90<A<208$, was shown to
occur in the low-mass asymptotic giant branch (AGB) stars during recurrent 
thermal pulses (Gallino et al. 1998; Busso et al. 1999). In particular, they 
showed that the main s-component is due to  
low metallicity ($[Fe/H] < - 1.5$) low mass AGB stars 
(1.5-3.0 $M_{\odot}$).\\
The s-process mechanism operating in the AGB model is dependent on the 
initial stellar 
metallicity. In fact, although the $^{13}C$ pocket, which acts as neutron producer, is of ``primary origin'' in 
the work of Gallino et al. (1998) and  Busso et al. (1999, 2001),
the ensuing s-process production is dependent on the initial abundance 
of the Fe-group seeds, i.e. on the stellar metallicity.
The neutron exposure (the neutron flux per nuclei seed) is indeed roughly proportional 
to the number of available neutron sources (the $^{13}C$ nuclei) 
per seed (the iron nuclei), hence inversely proportional to the stellar metallicity.

On the other hand, the weak s-component is responsible for the s-process nuclides up to  $A\simeq90$
and it is recognized as the result of neutron capture in advanced evolution in  massive 
stars(see Raiteri et al. 1993).
Finally, the strong-s component was introduced by \cite{b10} in order to reproduce more than 50\%
 of solar $^{208}Pb$.\\

The r-process takes place in extremely neutron-rich environments in which 
the mean time between two
successive neutron captures is very short, compared with the time necessary for the $\beta$-decay.
 Several scenarios have been proposed for the origin of r-process elements: neutrino winds in 
core-collapse supernovae (Woosley et al. 1994), the collapse of ONeMg cores resulting from stars
with initial masses in the range 8-10$M_{\odot}$ (Wanajo et al. 2003) and neutron star mergers 
(Freiburghaus et al. 1999),  even if this last scenario seems to be ruled out from recent
 work of Argast et al. (2004) at least as major responsible of r-process enrichment in our Galaxy.
 In any case, no clear consensus has been achieved and r-process nucleosynthesis 
remains still uncertain and until now, as far as we know, theoretical prescriptions for the r-process
 production still do not exist with the exception of the results of Wanajo et al. (2003) and
Woosley and Hoffmann (1992). 
However, the results of the model of Wanajo et al.  cannot be used in galactic chemical evolution models
 because, they do not take into account the fallback (after the SN explosion 
some material can fall back to the central collapsing neutron star) and so the amount
 of neutron capture elements produced 
is probably too high
(about 2 order of magnitude higher than the chemical evolution predictions).
Furthermore, Woosley and Hoffmann (1992) have given prescriptions for r-process only until $^{107}Ru$.
In order to shed light on the nature (s- and/or r- processes)
of heavy elements such as Ba and Eu one should examine the abundances of 
these elements in Galactic stars of all metallicities. These abundances can, 
in fact, give us clues to interpret their nucleosynthetic origin.
In the last few years a great deal of observational work for galactic stars 
appeared:  \citet{b26}, \citet{b330}, \citet{b500}, \citet{b12}, Mashonkina \& Gehren (2000, 2001),
 \citet{b190}, \citet{b170}, \citet{b172}.

One striking aspect of the data relative to both Ba 
and Eu is the large spread observed in the [Ba/Fe] and [Eu/Fe] ratios in halo 
stars (e.g. Mc William et al.1995; Ryan et al.1996).
Although this spread seems to be real, it is not found for the  
[$\alpha$/Fe] ratios in very metal poor stars (down to $[Fe/H]=-4.0$, 
Cayrel et al. 2004). This fact suggests that the spread, 
if real, is a characteristic of  the nuclear capture elements
and not only due to an inhomogeneous mixing in the early halo phases,
as suggested by several authors (Tsujimoto et al. 1999; Ishimaru \& Wanajo 1999).

Previous studies of the evolution of s- and r- process elements are from 
Mathews et al. (1992), Pagel \& Tautvaisiene (1997), Travaglio et al. (1999). 
In the Mathews et al. (1992) paper it was suggested that the observed 
apparent decrease of the abundance of Eu for [Fe/H] $<-2.5$ could be due to 
the fact that Eu originates mainly in low mass core-collapse SNe 
(7-8 $M_{\odot}$). 
Pagel \& Tautvaisiene (1997) suggested that to reproduce the observed 
behaviour of Ba it is necessary to assume that at early stages Ba is 
produced as an r-process element by a not well identified range of massive stars.
A similar conclusion was reached by Travaglio et al. (1999) who 
showed that the evolution of Ba cannot be explained by assuming that 
this element is only a s-process element mainly formed in stars with initial 
masses 2-4$M_{\odot}$, but an r-process origin for it should be considered. 
In fact, in the former hypothesis a very late appearance of Ba should be 
expected, at variance with the observations indicating that Ba was already 
produced at [Fe/H]=-4.0. They suggested that low mass SNII 
(from 8 to 10$M_{\odot}$) could be responsible for the r-component of Ba.
An attempt to explain  the observed spread in s- and r-elements 
can be found later in Tsujimoto et 
al. (1999) and Ishimaru \& Wanajo (1999),
who claim for an inefficient mixing in 
the early galactic phases and attribute the spread to the fact that 
we observe the pollution due to single supernovae.
Ishimaru \& Wanajo (1999) also concluded that the Eu should originate as an 
r-process element in stars with masses in the range 8-10$M_{\odot}$.
This latter suggestion was confirmed by Ishimaru et al. (2004) by comparing model
 predictions with new data from Subaru indicating subsolar [Eu/Fe] ratios in three 
stars very metal poor ([Fe/H]$<-3.0$).\\

In this paper we present the results of a chemical evolution model,
based on the original two-infall model of Chiappini et al. (1997) for the Milky Way 
in the latest version developed by Chiappini et al. (2003) and adopted in Fran\c cois et al. (2004).

We compare the predictions relative to the evolution of Ba and Eu 
with the newest data  of Fran\c cois et al. (2005).
A spread, although lower than in previous data, still exists 
for [Ba/Fe] and [Eu/Fe] at low metallicities.
We attempt to give an explanation of this spread without invoking 
only the inhomogeneous halo mixing.

The paper is organized as follows: in Sect. 2 we present the observational 
data, in Sect. 3 the chemical evolution model is presented and in Sect. 
4 the adopted nucleosynthesis prescriptions  are described.
In Sect. 5  we present the results and in Sect. 6 some conclusions are 
drawn.

\section{Observational data}\label{data}

In this paper, we preferentially used the most recent available data based on high quality spectra collected
with efficient spectrographs and 8-10 m class telescopes.
In particular, for the extremely  metal poor stars ([Fe/H] between $-4$ and $-3$), we adopted the recent
 results from UVES Large Program
"First Star'' (Cayrel  et al. 2004, Fran\c cois et al. 2005). This sample is made of 31 extremely metal-poor
halo stars selected in the HK survey (Beers et al. 1992, 1999). 
We can deduce from the kinematics of these stars that they were born at very different places in the 
Galactic halo.
This overcomes the possibility of a selection bias.
The analysis is made in a systematic and homogeneous way, from very high 
quality data, giving  abundance ratios of unprecedented  accuracy in this metallicity range.
For the abundances in the remaining range of [Fe/H], we took published high quality data in the
literature from various sources: \citet{b500}, \citet{b12},
 Mashonkina \& Gehren (2000, 2001),
 \citet{b190}, \citet{b170}, \citet{b172}.
 All of this data are relative to solar abundances of \citet{b17}.

\section{Chemical evolution model for the Milky Way}
We model the formation of the Galaxy assuming two main infall episodes: the first forms the halo
 and the thick disk, the second the thin disk. The timescale for the formation of 
the halo-thick disk is $\sim1Gyr$. The timescale 
for the thin disk is much longer, implying that the infalling gas
 forming the thin disk comes mainly from the
 intergalactic medium and not only from the halo (Chiappini et al. 1997). 
Moreover, the formation of the thin disk is assumed to be function of the galactocentric distance,
 leading to an inside out scenario for the Galaxy  disk built up (Matteucci \& Fran\c cois 1989).
 The main characteristic  of the two-infall model is an almost independent evolution
 between the halo and the thin disk
(see also Pagel \& Tautvaisienne 1995). A threshold in the star formation process (Kennicutt 1989, 1998,
 Martin \& Kennicutt 2001) is also adopted.
 The model well reproduces an extended set of observational constraints both for the solar
neighborhood and for the whole disc. One of the most important observational constraints
is represented by the various relations between the abundances of metals (C,N,O,$\alpha$-elements,
iron peak elements) as functions of the [Fe/H] abundance  (see Chiappini et al. 2003).
The equation below describes the time evolution  of $G_{i}$, namely the mass fraction of
 the element $i$ in the gas:

\begin{displaymath}
\dot{G_{i}}(t)=-\psi(r,t)X_{i}(r,t)
\end{displaymath}
\smallskip
\begin{displaymath}
+\int\limits^{M_{Bm}}_{M_{L}}\psi(t-\tau_{m})Q_{mi}(t-\tau_{m})\phi(m)dm
\end{displaymath}
\begin{displaymath}
+A\int\limits^{M_{BM}}_{M_{Bm}}\phi(M_{B})\cdot\left[\int\limits_{\mu_{m}}^{0.5}f(\mu)\psi(t-\tau_{m2})Q^{SNIa}_{mi}(t-\tau_{m2})d\mu\right]dM_{B}
\end{displaymath}
\begin{displaymath}
+(1-A)\int\limits^{M_{BM}}_{M_{Bm}}\psi(t-\tau_{m})Q_{mi}(t-\tau_{m})\phi(m)dm
\end{displaymath}
\begin{displaymath}
+\int\limits^{M_{U}}_{M_{BM}}\psi(t-\tau_{m})Q_{mi}(t-\tau_{m})\phi(m)dm
\end{displaymath}
\smallskip
\begin{equation}
+X_{A_{i}}A(r,t)
\end{equation}
where $X_{i}(r,t)$ is the abundance by mass of the element $i$ and $Q_{mi}$ indicates
the fraction of mass restored by star of mass $m$ in form of the element $i$, the so-called
``production matrix'' as originally defined by Talbot and Arnett (1973). We indicate with $M_{L}$ the lightest
mass which contributes to the chemical enrichment and it is set at $0.8M_{\odot}$; the upper mass limit, $M_{U}$, 
is set at $100M_{\odot}$. 

\begin{figure}
\begin{center}
\includegraphics[width=0.5\textwidth]{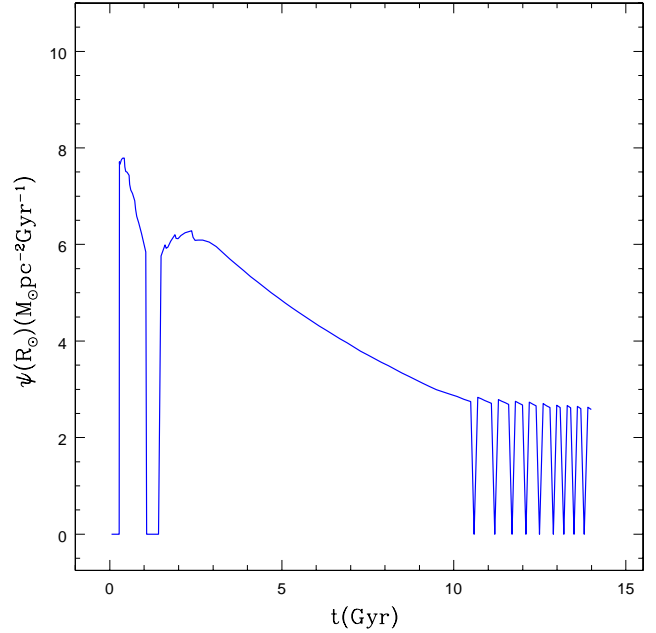}
\caption{ The SFR expressed  in $M_{\odot}pc^{-2}Gyr^{-1}$ as predicted by the two infall model. The 
gap in the SFR at the end of the halo-thick disc phase is evident. The oscillations are due to the fact
that at the late times in the galactic disc the surface gas density is always close to the threshold density.}
\label{SFR}
\end{center}
\end{figure}

The star formation rate (SFR) $\psi(r,t)$ is defined:
\begin{equation}
\psi(r,t)=\nu\left(\frac{\Sigma(r,t)}{\Sigma(r_{\odot},t)}\right)^{2(k-1)}
\left(\frac{\Sigma(r,t_{Gal})}{\Sigma(r,t)}\right)^{k-1}G^{k}_{gas}(r,t)
\end{equation}
$\nu$ is the efficiency of the star formation process and is set to be $2Gyr^{-1}$
for the Galactic halo ($t<1Gyr$) and $1Gyr^{-1}$ for the disk ($t\ge1Gyr$).
$\Sigma(r,t)$ is the total
surface mass density, $\Sigma(r_{\odot},t)$ the total surface mass density at the 
solar position, $G_{gas}(r,t)$ the surface density normalized to the present time
total surface mass density in the disk $\Sigma_{D}(r,t_{Gal})$, where $t_{Gal}=14Gyr$ is the age 
assumed for the Milky Way and $r_{\odot}=8kpc$ the solar galactocentric distance 
(Reid 1993). The gas surface exponent, $k$, is set equal to 1.5.
With these values for the parameters the observational constraints, in particular in the solar vicinity,
are well fitted.
Below a critical threshold of the gas surface density ($7M_{\odot}pc^{-2}$) we assume no star formation.
This naturally produces a hiatus in the SFR between the halo-thick disk phase and the thin disk phase.
In Fig. (\ref{SFR}) is shown the predicted star formation rate for the halo-thick disc phase
and the thin disc phase, respectively.

For $\phi$, the initial mass function (IMF), we use the \citet{b37}, constant in time and space.
$\tau_{m}$ is the evolutionary lifetime of stars as a function of their mass ``m''.

\begin{figure}
\begin{center}
\includegraphics[width=0.5\textwidth]{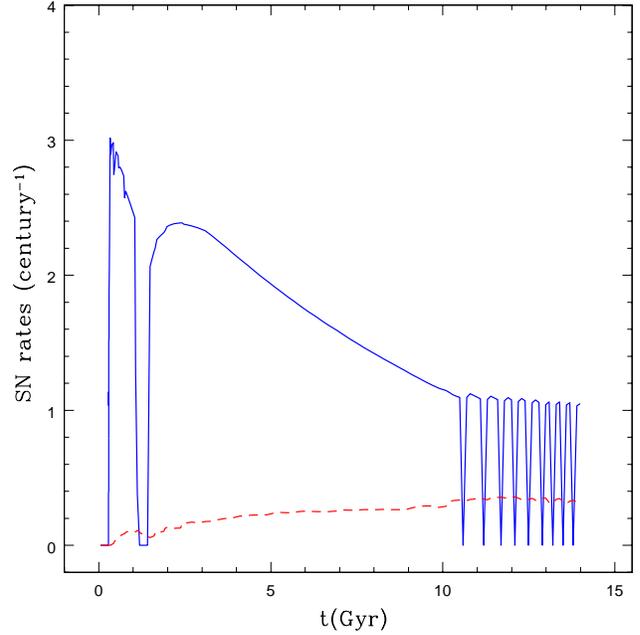}
\caption{Predicted SN II (continuous line) and Ia (dashed line) rates by the two infall model.}
\label{SNII}
\end{center}
\end{figure}

The SNeIa rate has been computed following \citet{b16} and \citet{b24} and it is expressed as:
\begin{equation}
R_{SNeIa}=A\int\limits^{M_{BM}}_{M_{Bm}}\phi(M_{B})(\int\limits^{0.5}_{\mu_{m}}f(\mu)\psi(t-\tau_{M_{2}})d\mu) dM_{B}
\end{equation}
where $M_{2}$ is the mass of the secondary, $M_{B}$ is the total mass of the binary
system, $\mu=M_{2}/M_{B}$, $\mu_{m}=max\left[M(t)_{2}/M_{B},(M_{B}-0.5M_{BM})/M_{B}\right]$, 
$M_{Bm}= 3 M_{\odot}$, $M_{BM}= 16 M_{\odot}$. The IMF is represented by $\phi(M_{B})$
and refers to the total mass of the binary sistem for the computation of the SNeIa rate,
$f(\mu)$ is the distribution function for the mass fraction of the secondary:
\begin{equation}
f(\mu)=2^{1+\gamma}(1+\gamma)\mu^{\gamma}  
\end{equation}
with $\gamma=2$; A is the fraction of systems in the appropriate mass range, which can give rise
to SNeIa events. This quantity is fixed to 0.05 by reproducing the observed SNeIa rate at
the present time (Cappellaro et al. 1999). Note that  in the case of SNIa the``production matrix''
is indicated with  $Q^{SNIa}_{mi}$ because of its different nucleosynthesis contribution
(for details refer to Matteucci and Greggio 1986).
In Fig. \ref{SNII} we show the predicted type II and Ia SN rates. The type II SN rate follows
the SFR, as expected, whereas the type Ia SN rate does not have this feature due to the nature
of type Ia SN progenitors, which are assumed to be low-intermediate mass star with a long 
evolutionary time scales.

The last term in equation 1 represents the accretion and it is defined as:
\begin{equation}
A(r,t)=a(r)e^{-t/\tau_{H}}+b(r)e^{(t-t_{max})/\tau_{D}(r)}
\end{equation}
$X_{A_{i}}$ are the abundances of infalling material, assumed primordial, $t_{max}=1Gyr$
is the time for the maximum infall rate on the thin disk, $\tau_{H}=2.0Gyr$ is the time scale
for the formation of the halo thick-disk and $\tau_{D}$ is the timescale of the thin disk,
 function of the galactocentric distance:
\begin{equation}
\tau_{D}=1.033r(kpc)-1.267Gyr
\end{equation}
The coefficients $a(r)$ and $b(r)$ are constrained by the present day
 total surface mass density as a function of galactocentric distance.
In particular, $b(r)$ is assumed to be different from zero only for $t>t_{max}$,
 where $t_{max}$ is the time of maximum infall on the thin disc (see Chiappini et al. 2003, for details).

\section{Nucleosynthesis Prescriptions}{\label{NP}}
\subsection{S-process}{\label{NP_BaS}}

\begin{figure}
\begin{center}
\includegraphics[width=0.5\textwidth]{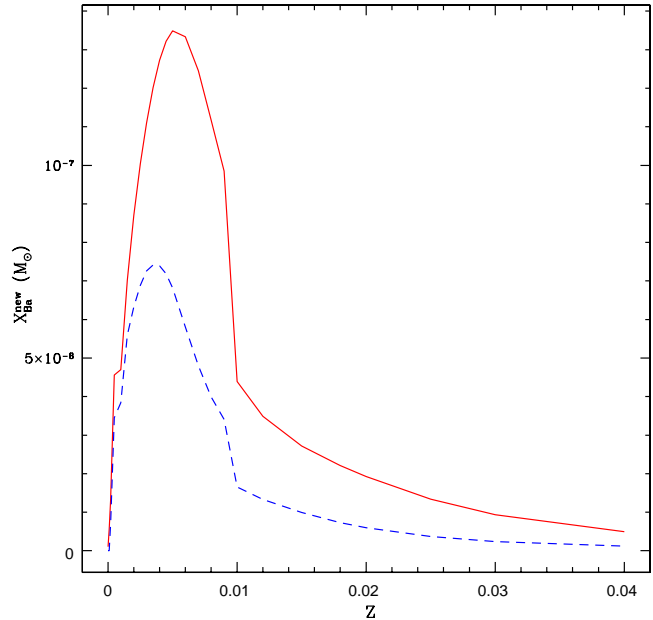}
\caption{The stellar yields  $X^{new}_{Ba}$ from the paper of \citet{b4} plotted versus metallicity.
 In dashed
line the prescriptions for stars of $1.5M_{\odot}$, in solid line for stars of $3M_{\odot}$.} \label{sBa2}
\end{center}
\end{figure}

 We have adopted the yields of Busso et al. (2001) in the mass
 range 1.5-3$M_{\odot}$ for the s-main component. 
In this process, the dependence on the metallicity is very important. In fact,
the s-process elements are made by accretion of neutrons on seed elements
(in particular iron) already present in the star. Therefore, this Ba 
component behaves like a secondary element.
The neutron flux is due to the reaction $^{13}C(\alpha,n)^{16}O$ which
can easily be activated at the low temperature of these stars
(see Busso et al. 1999).
The yields are shown in Table \ref{SBa}
and Fig. \ref{sBa2} as functions of the initial metallicity of the stars.
The theoretical results by Busso et al. (2001) suggest 
negligible Europium production  
in the s-process and therefore we neglected this component in our work.
We have added for models 1 and 2 (see table \ref{model}) an extension to the
theoretical result of Busso et al. (2001)  in the mass range $1-1.5M_{\odot}$
by simply scaling with the mass the values obtained for stars of $1.5 M_{\odot}$.
We have extended the prescription in order to better fit the data with a [Fe/H]
supersolar and it does not change the results of the model at $[Fe/H]<0$.

\begin{table}

\caption{The stellar yields in the range $1.5-3M_{\odot}$ from the paper of \citet{b4}.}\label{SBa}

\begin{center}
\begin{tabular}{|c|c|c|}
\hline

$ Metallicity $ & $X^{new}_{Ba}$ for $1.5M_{\odot}$& $X^{new}_{Ba}$ for $3M_{\odot}$  \\

\hline\hline

 0.20$\cdot10^{-3}$ &  0.69$\cdot10^{-8}$  & 0.13$\cdot10^{-7}$ \\
 0.10$\cdot10^{-2}$ &  0.38$\cdot10^{-7}$  & 0.46$\cdot10^{-7}$ \\
 0.20$\cdot10^{-2}$ &  0.63$\cdot10^{-7}$  & 0.87$\cdot10^{-7}$ \\
 0.30$\cdot10^{-2}$ &  0.72$\cdot10^{-7}$  & 0.11$\cdot10^{-6}$ \\
 0.40$\cdot10^{-2}$ &  0.73$\cdot10^{-7}$  & 0.12$\cdot10^{-6}$ \\
 0.50$\cdot10^{-2}$ &  0.68$\cdot10^{-7}$  & 0.13$\cdot10^{-6}$ \\
 0.60$\cdot10^{-2}$ &  0.58$\cdot10^{-7}$  & 0.13$\cdot10^{-6}$ \\ 
 0.70$\cdot10^{-2}$ &  0.47$\cdot10^{-7}$  & 0.12$\cdot10^{-6}$ \\
 0.80$\cdot10^{-2}$ &  0.39$\cdot10^{-7}$  & 0.11$\cdot10^{-6}$ \\
 0.90$\cdot10^{-2}$ &  0.34$\cdot10^{-7}$  & 0.98$\cdot10^{-7}$ \\
 0.10$\cdot10^{-1}$ &  0.16$\cdot10^{-7}$  & 0.43$\cdot10^{-7}$ \\ 
 0.11$\cdot10^{-1}$ &  0.14$\cdot10^{-7}$  & 0.39$\cdot10^{-7}$ \\ 
 0.12$\cdot10^{-1}$ &  0.13$\cdot10^{-7}$  & 0.34$\cdot10^{-7}$  \\
 0.13$\cdot10^{-1}$ &  0.12$\cdot10^{-7}$  & 0.32$\cdot10^{-7}$  \\
 0.14$\cdot10^{-1}$ &  0.11$\cdot10^{-7}$  & 0.29$\cdot10^{-7}$  \\
 0.15$\cdot10^{-1}$ &  0.99$\cdot10^{-8}$  & 0.27$\cdot10^{-7}$  \\
 0.16$\cdot10^{-1}$ &  0.90$\cdot10^{-8}$  & 0.25$\cdot10^{-7}$  \\
 0.17$\cdot10^{-1}$ &  0.81$\cdot10^{-8}$  & 0.23$\cdot10^{-7}$  \\
 0.18$\cdot10^{-1}$ &  0.73$\cdot10^{-8}$  & 0.22$\cdot10^{-7}$  \\
 0.19$\cdot10^{-1}$ &  0.66$\cdot10^{-8}$  & 0.20$\cdot10^{-7}$  \\
 0.20$\cdot10^{-1}$ &  0.59$\cdot10^{-8}$  & 0.19$\cdot10^{-7}$  \\   
 0.30$\cdot10^{-1}$ &  0.24$\cdot10^{-8}$  & 0.94$\cdot10^{-8}$  \\   
 0.40$\cdot10^{-1}$ &  0.12$\cdot10^{-8}$  & 0.50$\cdot10^{-8}$  \\

\hline\hline                                                    

\end{tabular}

\end{center}

\end{table}

\subsection{R-process}{\label{NP_BaR}}
As already said in the introduction the production of r-process elements is still 
a challenge
for astrophysics and even for nuclear physics, due to the fact that 
the nuclear properties of thousand of nuclei located between 
the valley of $\beta$ stability and the neutron drip line, necessary to
correctly  compute this process, are ignored.
In our models we 
have tested 6 different nucleosynthesis prescriptions for the r-process Ba and Eu, 
as shown in tables  \ref{model}, \ref{rBa} and \ref{rBa2}.
Some of the prescriptions refer to models by Travaglio et al. (2001) (model 3)
and Ishimaru et al. (2004)(models 4, 5 and 6), whereas the others contain yields chosen 
rather ``ad hoc''.

\begin{figure}
\begin{center}
\includegraphics[width=0.5\textwidth]{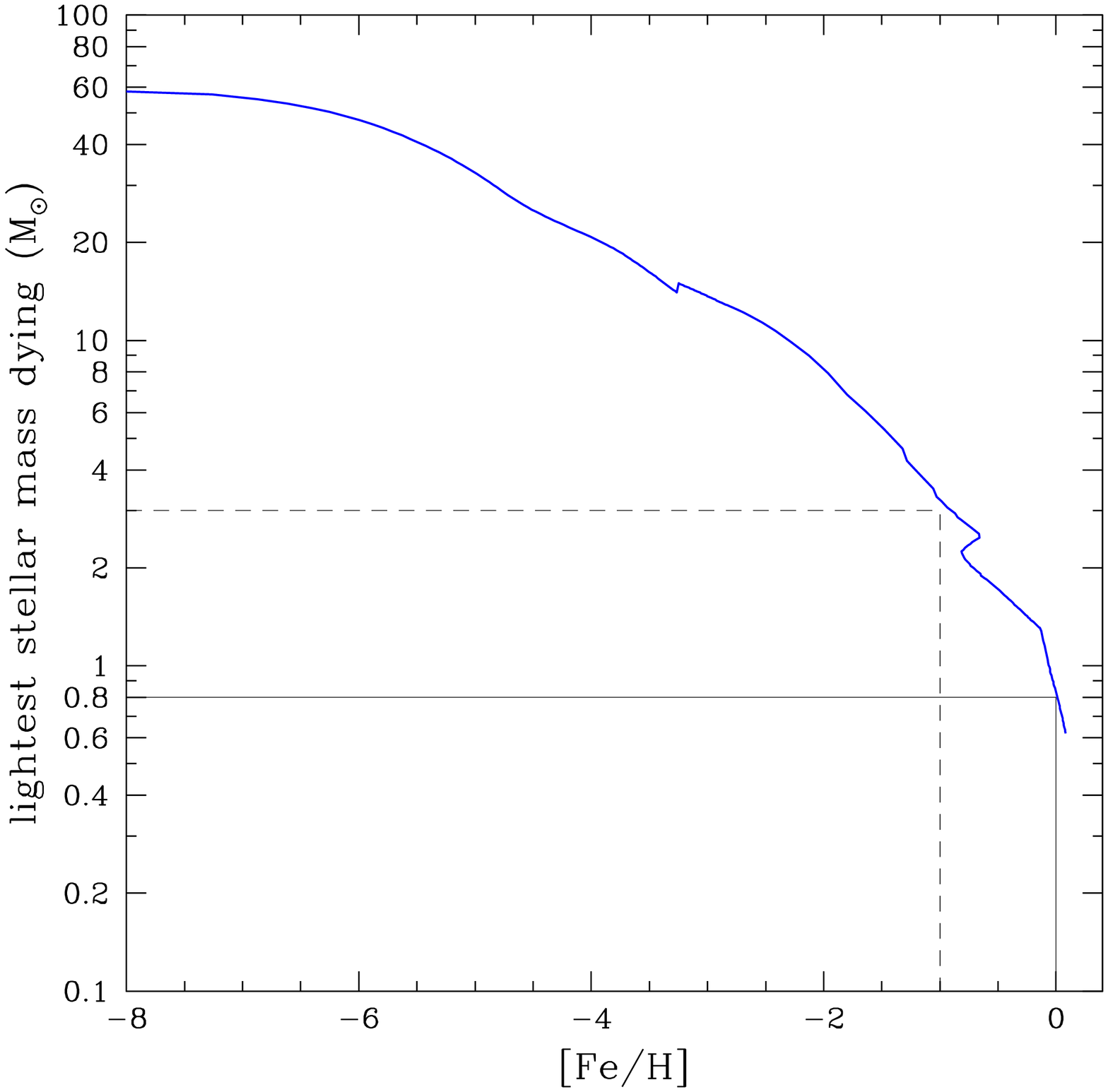}
\caption{In the plot we show the lightest stellar mass dying at the time corresponding to a given [Fe/H].
The solid line indicates the solar abundance ([Fe/H]=0), corrisponding to a lightest dying mass star of $0.8M_{\odot}$,
the dashed line indicates the [Fe/H]=-1 corrisponding to a lightest dying star mass of $3M_{\odot}$.}
\label{MORTE}
\end{center}
\end{figure}

In the case of Ba we have included an r-process component,
produced in massive stars in the range 12-30$M_{\odot}$ in  model 1
and in the range 10-25$M_{\odot}$ in model 2.
In Fig. \ref{MORTE} we show the lightest stellar mass dying as function of
the ratio [Fe/H] in our chemical evolution model; it is clear 
from this plot that  it is impossible to explain the 
observed abundances of [Ba/Fe] in stars with $[Fe/H]<-2$ without the component produced
in massive stars. In fact, the first stars,
which produce s-processed Ba (see Sect. \ref{NP_BaS}), have a mass of $3M_{\odot}$ 
and they start to enrich the ISM only at $[Fe/H] \ge -2$.

We stress that Travaglio et al. (2001) predicted r-process Ba only from 
stars in the range 8-10$M_{\odot}$, but their conclusions were based upon 
an older set of observational data.

Moreover, we considered another independent indication for the r-process 
production of Barium; 
Mazzali and Chugai (1995) observed Ba lines 
in SN 1987A, which  had a progenitor star of 20$M_{\odot}$.
These lines of Ba are well reproduced with a overabundance
factor $f=X_{obs}/X_{i}=5$ (typical metal abundance for LMC $X_{i}=(1/2.75)\times$ solar).
 From this observational data we can derive
a $X_{Ba}^{new}\sim 2 \cdot 10^{-8}$, which is in agreement with our 
prescriptions.

For the Eu  we assumed that it is completely due to the r-process
and that the yields originate from massive stars  in the range 
12-30$M_{\odot}$ in model 1 and 10-25$M_{\odot}$ in model 2,
 as shown in table \ref{model}.

In particular, our  choice is done
with the purpose of best fitting  the plots [Ba/Fe] vs.[Fe/H], 
[Eu/Fe] vs.[Fe/H] and [Ba/Eu] vs [Fe/H] as well as the Ba and Eu
 solar abundance (taking in account the contribution of the
 low-intermediate mass star in case of the Ba). 

We have tested prescriptions for Ba and Eu both for a primary production
 and a secondary production (with a dependence on the metallicity).
In the first case the main feature of the  yields
is a big enhancement in the mass range $12-15M_{\odot}$
(model 1) with no dependence on the metallicity and so
the elements are considered as primary elements.
In the case of metallicity dependence (model 2), the yield behaviour
is chosen to have a strong enhancement in the range of metallicity $5\cdot10^{-7}<Z<1\cdot10^{-5}$
 with almost constant yield for Eu and Ba in the whole mass range for
a given metallicity.

\subsection{Iron}
For the nucleosynthesis prescriptions of the Fe, we adopted those suggested in Fran\c cois 
et al. (2004), in particular  we considered the yields of Woosley \& Weaver (1995) (hereafter WW95) for a solar chemical composition. We 
remind that the yields suggested for several elements by Fran\c cois et al. (2004) are those reproducing at best the observed 
[X/Fe] vs. [Fe/H] at all metallicities in the solar vicinity.

\begin{table*}

\caption{Model parameters. The yields $X^{new}_{Ba}$ are expressed as mass fractions. The subscript  
``ext'' stands for extended (the yields have been extrapolated down to $1M_{\odot}$) and $M_{*}$ for
 the mass of the star.}\label{model}

\begin{tabular}{|c|c|c|c|c|}
\hline\hline

Mod             &  s-process Ba          &  r-process  Ba   & s-process Eu & r-process Eu \\
\hline\hline
1               & $1.-3M_{\odot}$        & $12-30M_{\odot}$ &   none            & $12-30M_{\odot}$ \\  
                & Busso et al.(2001)ext.    & yields table 3   &                   &  yields table 3  \\
\hline\hline
2               &  $1.-3M_{\odot}$        & $10-25M_{\odot}$ &   none            & $10-25M_{\odot}$ \\  
                &  Busso et al.(2001)ext. & yields table 4   &                   &  yields table 4  \\
\hline 
3               &  $1.5-3M_{\odot}$      & $8-10M_{\odot}$  &   none            & $12-30M_{\odot}$ \\  
                &  Busso et al.(2001)    & $X^{new}_{Ba}=5.7\cdot10^{-6}/M_{*}$& &  yields table 3 \\
                &                        & (Travaglio et al. 2001)             & &  \\

\hline\hline
4               &  $1.5-3M_{\odot}$      & $10-30M_{\odot}$ &   none            &  $8-10M_{\odot}$\\  
                &  Busso et al.(2001)    & yields table 3   &                   &  $X^{new}_{Eu}=3.1\cdot10^{-7}/M_{*}$ \\
                &                        &                  &                   &  (Ishimaru et al.2004 Mod.A)\\
\hline
5               &  $1.5-3M_{\odot}$      & $10-30M_{\odot}$ &   none            &  $20-25M_{\odot}$\\  
                &  Busso et al.(2001)    & yields table 3   &                   &  $X^{new}_{Eu}=1.1\cdot10^{-6}/M_{*}$ \\
                &                        &                  &                   &  (Ishimaru et al.2004 Mod.B)\\
\hline
6               &  $1.5-3M_{\odot}$      & $10-30M_{\odot}$ &   none            &  $>30M_{\odot}$\\  
                &  Busso et al.(2001)    & yields table 3   &                   &  $X^{new}_{Eu}=7.8\cdot10^{-7}/M_{*}$ \\
                &                        &                  &                   &  (Ishimaru et al.2004 Mod.C)\\

\hline\hline
\end{tabular}
\end{table*}

\begin{table*}

\caption{The stellar yields for Barium and Europium in massive stars (r-process)
in the case of a primary origin.} \label{rBa}

\centering
\begin{minipage}{90mm}

\begin{tabular}{|c|c|c|}
\hline

$M_{star}$  & $ X_{Ba}^{new}$  & $ X_{Eu}^{new}$\\

\hline\hline

12.   & 9.00$\cdot10^{-7}$ &  4.50$\cdot10^{-8}$  \\ 
15.   & 3.00$\cdot10^{-8}$ &  3.00$\cdot10^{-9}$ \\   
30.   & 1.00$\cdot10^{-9}$ &  5.00$\cdot10^{-10}$ \\

\hline\hline

\end{tabular}

\end{minipage}

\end{table*}

\begin{table*}

\caption{The stellar yields for Ba and Eu in massive stars (r-process)
in the case of secondary origin. The mass fraction does not change 
in function of the stellar mass.} \label{rBa2}

\centering

\begin{minipage}{90mm}

\begin{tabular}{|c|c|c|c|}
\hline

$Z_{star}$  & $X_{Ba}^{new}$   & $ X_{Eu}^{new}$ \\
            & $10-25M_{\odot}$ & $10-25M_{\odot}$ \\
\hline\hline

           $Z<5\cdot10^{-7}$.   &  1.00$\cdot10^{-8}$ &  5.00$\cdot10^{-10}$  \\ 
$5\cdot10^{-7}<Z<1\cdot10^{-5}$ &  1.00$\cdot10^{-6}$ &  5.00$\cdot10^{-8}$  \\   
           $Z>1\cdot10^{-5}$    &  1.60$\cdot10^{-7}$ &  8.00$\cdot10^{-9}$  \\

\hline\hline

\end{tabular}

\end{minipage}

\end{table*}

\section{Results}

\subsection{Trends}\label{trends}

In this Sect. we investigate how the different models fit the 
the trends of the abundances ratios for [Ba/Fe],
[Eu/Fe] and [Ba/Eu] versus [Fe/H] and even for [Ba/Eu] versus [Ba/H].

To better investigate the trends of the data we have decided to divide
in several bins the [Fe/H] axis and the [Ba/H] axis and compute
the mean and the standard deviation from the mean of the ratios 
[Ba/Fe], [Eu/Fe] and [Ba/Eu] for  all the data inside each bin.

In table \ref{meanBa} we show the results of this computation
for [Ba/Fe] versus [Fe/H], in table \ref{meanEu}
for [Eu/Fe] and [Ba/Eu] versus [Fe/H] and finally in table \ref{meanBaH}
for [Ba/Eu] versus [Ba/H].

 Obviously having the ranges [Ba/H] and [Fe/H] different, we have bins
of different width. 

We have divided in a different way the [Fe/H] for [Ba/Fe] 
ratio and the [Fe/H] for [Eu/Fe] and [Ba/Eu] ratios because
the [Eu/Fe] ratio for 12 stars at very low
metallicity is only an upper limit and therefore the data of these stars
have not been considered in the computation 
of the mean and the standard deviation for [Eu/Fe] and [Ba/Eu] ratios.

In the case of [Ba/Eu] and [Eu/Fe] we have simply divided the
[Fe/H] axis in 15 bins of equal dimension (see table \ref{meanEu}); for [Ba/Fe]
 we have divided in 18 bins the [Fe/H] but 
we have merged the first three bins (starting from the lowest value in [Fe/H])
 in a single bin in order to have enough data in the first bin  (see table \ref{meanBa}).
Finally for [Ba/Eu] versus [Ba/H] we have splitted in 16 equal bins but
again we have merged the first two pairs in two bins for the same reason 
 (see table \ref{meanBaH}).

\begin{table*}

\caption{Results after the computation of the mean for the data inside bins
 along the [Fe/H] axis for the values of [Ba/Fe].} \label{meanBa}

\begin{tabular}{|c|c|c|c|c|}
\hline

bin center [Fe/H]& bin dim.[Fe/H]  & mean [Ba/Fe] & SD [Ba/Fe] &  N. of data in the bin \\
\hline\hline

 -3.82 & 0.75 & -1.25 &  0.30 &   6\\
 -3.32 & 0.25 & -0.96 &  0.50 &   7\\
 -3.07 & 0.25 & -0.65 &  0.65 &  11\\
 -2.82 & 0.25 & -0.37 &  0.60 &  17\\
 -2.57 & 0.25 & -0.15 &  0.40 &  11\\
 -2.32 & 0.25 &  0.09 &  0.58 &  13\\
 -2.07 & 0.25 &  0.23 &  0.50 &  15\\
 -1.82 & 0.25 &  0.10 &  0.20 &  20\\
 -1.58 & 0.25 &  0.08 &  0.15 &  27\\
 -1.33 & 0.25 &  0.20 &  0.22 &  16\\
 -1.08 & 0.25 &  0.07 &  0.19 &  20\\
 -0.83 & 0.25 & -0.03 &  0.08 &  30\\
 -0.58 & 0.25 & -0.04 &  0.14 &  59\\
 -0.33 & 0.25 &  0.05 &  0.20 &  46\\
 -0.08 & 0.25 &  0.03 &  0.13 &  53\\
  0.17 & 0.25 & -0.01 &  0.11 &  26\\

\hline\hline

\end{tabular}

\end{table*}

\begin{table*}

\caption{Results after the computation of the mean for the data inside bins
 along the [Fe/H] axis for the values of [Eu/Fe] and [Ba/Eu].} \label{meanEu}

\begin{tabular}{|c|c|c|c|c|c|c|}
\hline

bin center [Fe/H]& bin dim.[Fe/H]& mean [Eu/Fe]& SD [Eu/Fe]& mean [Ba/Eu]&SD [Ba/Eu]& N of data in the bin \\
\hline\hline
 -3.22 & 0.24 &  -0.10 &  0.21 & -0.71  & 0.25  &  5\\
 -2.98 & 0.24 &   0.08 &  0.60 & -0.57  & 0.13  & 12\\
 -2.74 & 0.24 &   0.46 &  0.60 & -0.64  & 0.11  & 14\\
 -2.49 & 0.24 &   0.45 &  0.28 & -0.52  & 0.17  &  7\\
 -2.25 & 0.24 &   0.38 &  0.36 & -0.38  & 0.33  & 11\\
 -2.01 & 0.24 &   0.51 &  0.34 & -0.36  & 0.26  & 10\\
 -1.77 & 0.24 &   0.29 &  0.22 & -0.20  & 0.19  & 19\\
 -1.53 & 0.24 &   0.44 &  0.15 & -0.39  & 0.22  & 21\\
 -1.28 & 0.24 &   0.42 &  0.20 & -0.26  & 0.31  & 18\\
 -1.04 & 0.24 &   0.39 &  0.13 & -0.38  & 0.15  & 16\\
 -0.80 & 0.24 &   0.32 &  0.12 & -0.35  & 0.14  & 36\\
 -0.56 & 0.24 &   0.23 &  0.14 & -0.27  & 0.20  & 55\\
 -0.32 & 0.24 &   0.18 &  0.10 & -0.13  & 0.23  & 44\\
 -0.07 & 0.24 &   0.04 &  0.07 & -0.02  & 0.14  & 51\\
  0.17 & 0.24 &  -0.02 &  0.07 &  0.00  & 0.12  & 26\\

\hline\hline

\end{tabular}

\end{table*}

\begin{table*}

\caption{Results after the computation of the mean for the data inside bins
 along the [Ba/H] axis for the values of [Ba/Eu].} \label{meanBaH}

\begin{tabular}{|c|c|c|c|c|}
\hline

bin center [Ba/H]& bin dim.[Ba/H]  & mean [Ba/Eu] & SD [Ba/Eu] &  N of data in the bin \\
\hline\hline

 -4.35 & 0.58 &  -0.75 & 0.26 &  4\\
 -3.76 & 0.58 &  -0.60 & 0.14 & 12\\
 -3.32 & 0.29 &  -0.55 & 0.14 &  3\\
 -3.02 & 0.29 &  -0.62 & 0.13 &  4\\
 -2.73 & 0.29 &  -0.58 & 0.24 & 13\\
 -2.43 & 0.29 &  -0.58 & 0.21 &  4\\
 -2.14 & 0.29 &  -0.44 & 0.13 &  7\\
 -1.84 & 0.29 &  -0.33 & 0.28 & 20\\
 -1.54 & 0.29 &  -0.33 & 0.20 & 25\\
 -1.25 & 0.29 &  -0.39 & 0.19 & 21\\
 -0.95 & 0.29 &  -0.31 & 0.20 & 36\\
 -0.66 & 0.29 &  -0.33 & 0.18 & 64\\
 -0.36 & 0.29 &  -0.13 & 0.14 & 43\\
 -0.07 & 0.29 &  -0.03 & 0.09 & 68\\

\hline\hline

\end{tabular}

\end{table*}

\begin{figure}
\begin{center}
\includegraphics[width=0.5\textwidth]{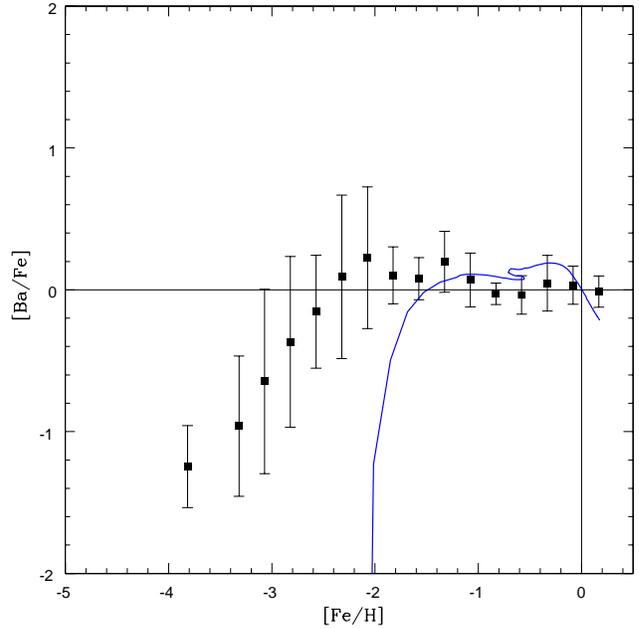}
\caption{In this Fig. is  plotted the [Ba/Fe] versus [Fe/H]. The squares are
  the mean values of the data bins described 
in the table \ref{meanBa}. As error bars we consider the standard deviation (see table \ref{meanBa}).
The solid line is the results of the model 3 from  table \ref{model}.}\label{trava}
\end{center}
\end{figure}

\begin{figure}
\begin{center}
\includegraphics[width=0.5\textwidth]{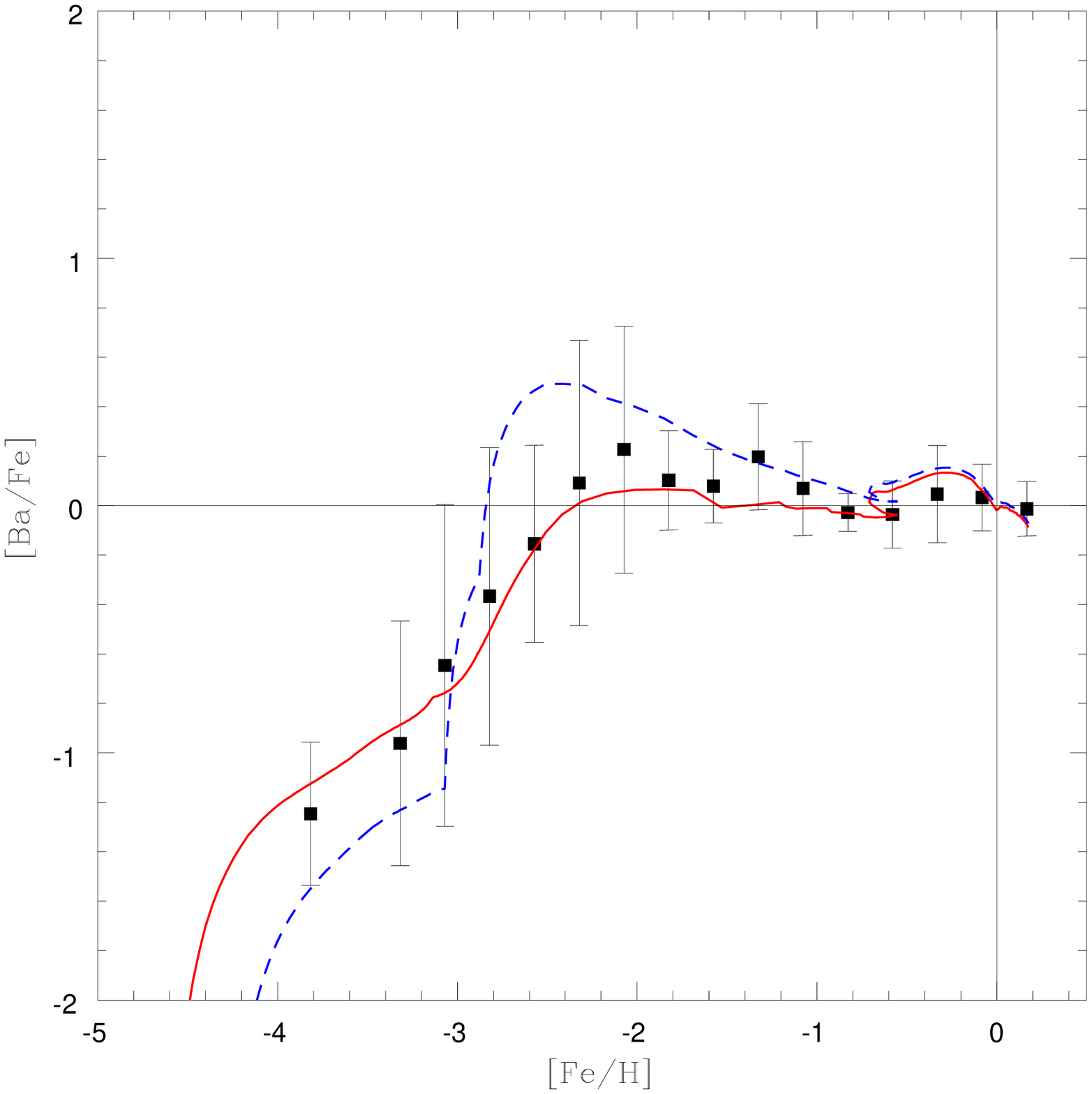}
\caption{The data are the same as in  Fig. \ref{trava}. In
 this Fig. we show in solid line the model 1 and in dashed line the model 2
 (models  are described in table \ref{model}) predictions.}\label{best1}
\end{center}
\end{figure}

In Fig. \ref{trava} we show the results for the model 3 (with
the yields used in Travaglio et al. 2001) for [Ba/Fe] versus [Fe/H].
As evident from Fig. \ref{trava}, this model does not fit the data.

Moreover, the model in Fig. \ref{trava} is different from the similar
model computed by Travaglio et al. (1999). 
In fact, we want to underline that we are using a different chemical
evolution model and this gives rise to different results.
The main difference between the two chemical evolution models
(the one of Travaglio and the present one) consists in the age-[Fe/H] relation
which grows more slowly in the model of Travaglio.
The cause for this difference is probably due to the different adopted
stellar lifetimes, to the different $M_{up}$ (i.e. the mass of the most massive star ending  its life
as C-O white dwarf) and to the yield  prescriptions for the iron  which
are probably the WW95 metallicity dependent ones in the model of Travaglio et al.(1999),
whereas we use the WW95 yields for the solar chemical compositios,
 which produce faster rise of Iron.

In fact, in order to better fit the new data
we have to extend the mass range for the production of the r-processed
Barium toward higher mass in order to reproduce [Ba/Fe] at lower metallicity.

In Fig. \ref{best1}, where we have plotted the 
predictions of the model 1 and the model 2  for [Ba/Fe] versus [Fe/H], it is clear that 
these models better fit the trend of the data.
In these models the upper mass limit for the production
of the r-processed Ba is 30$M_{\odot}$ in the case of model 1, and
25$M_{\odot}$ in the case of model 2.
However, the model 2 does not fit as well as model 1 the trend of the data
but we would like to stress that the prescriptions of model 2 are very simplistic.
In fact, there is no dependence on mass for a given metallicity
in the yields of Ba and Eu. This prescription is clearly an oversimplification
but our goal is to show how a model with yields only dependent on metallicity
works and so  we are able to better understand at least roughly 
if it could be a good approach or not.

\begin{figure}
\begin{center}
\includegraphics[width=0.5\textwidth]{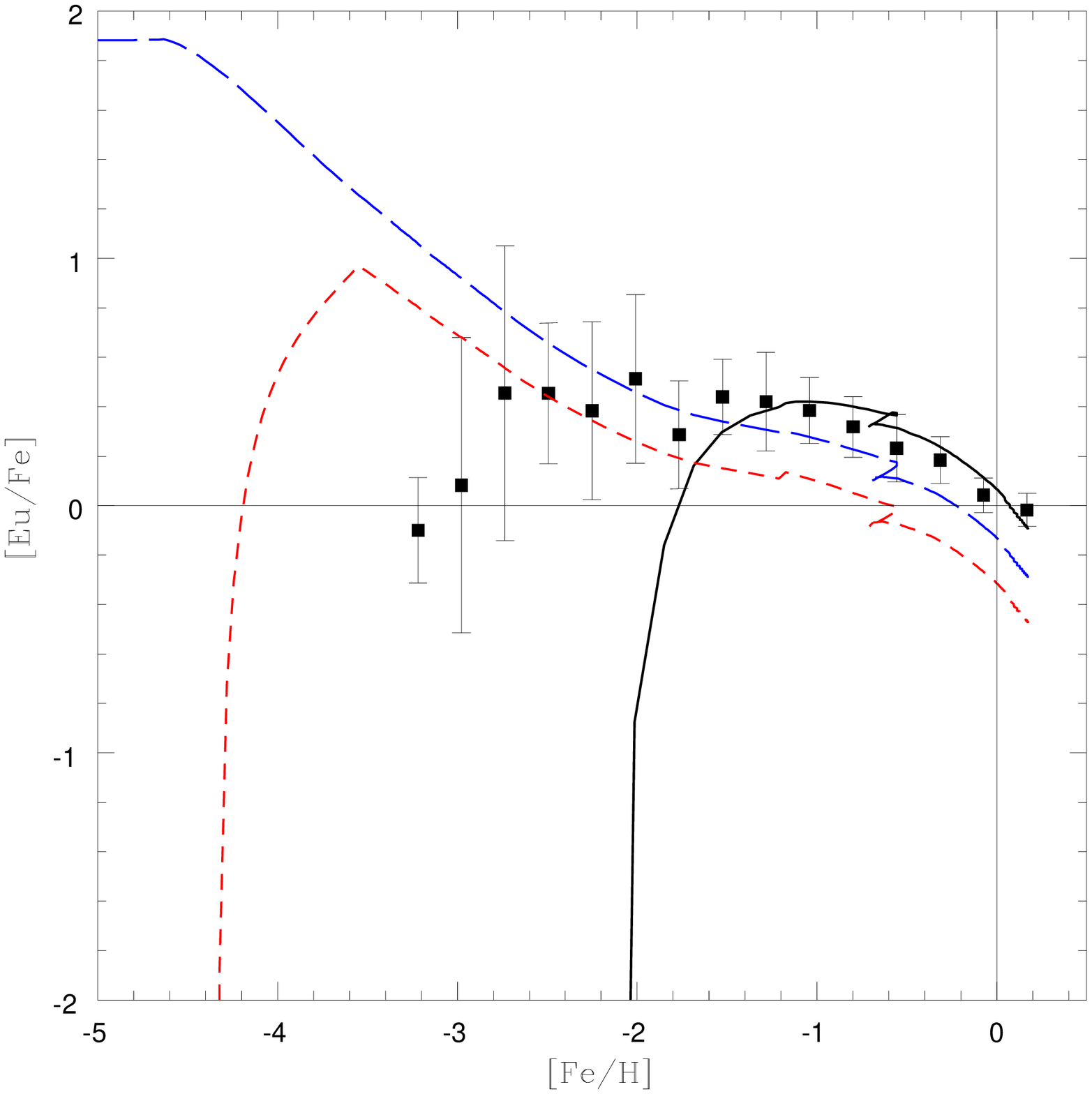}
\caption{In  graph is  plotted the [Eu/Fe] versus [Fe/H]. The squares are
  the mean values of the data bins described 
in the table \ref{meanEu}. As error bars we consider the standard deviation (see table \ref{meanEu}).
We show in solid line the results of model 4, in short dashed line the results of model 5
and in long dashed line the ones of model 6 (models  are described in table \ref{model})}\label{ISHI}
\end{center}
\end{figure}

\begin{figure}
\begin{center}
\includegraphics[width=0.5\textwidth]{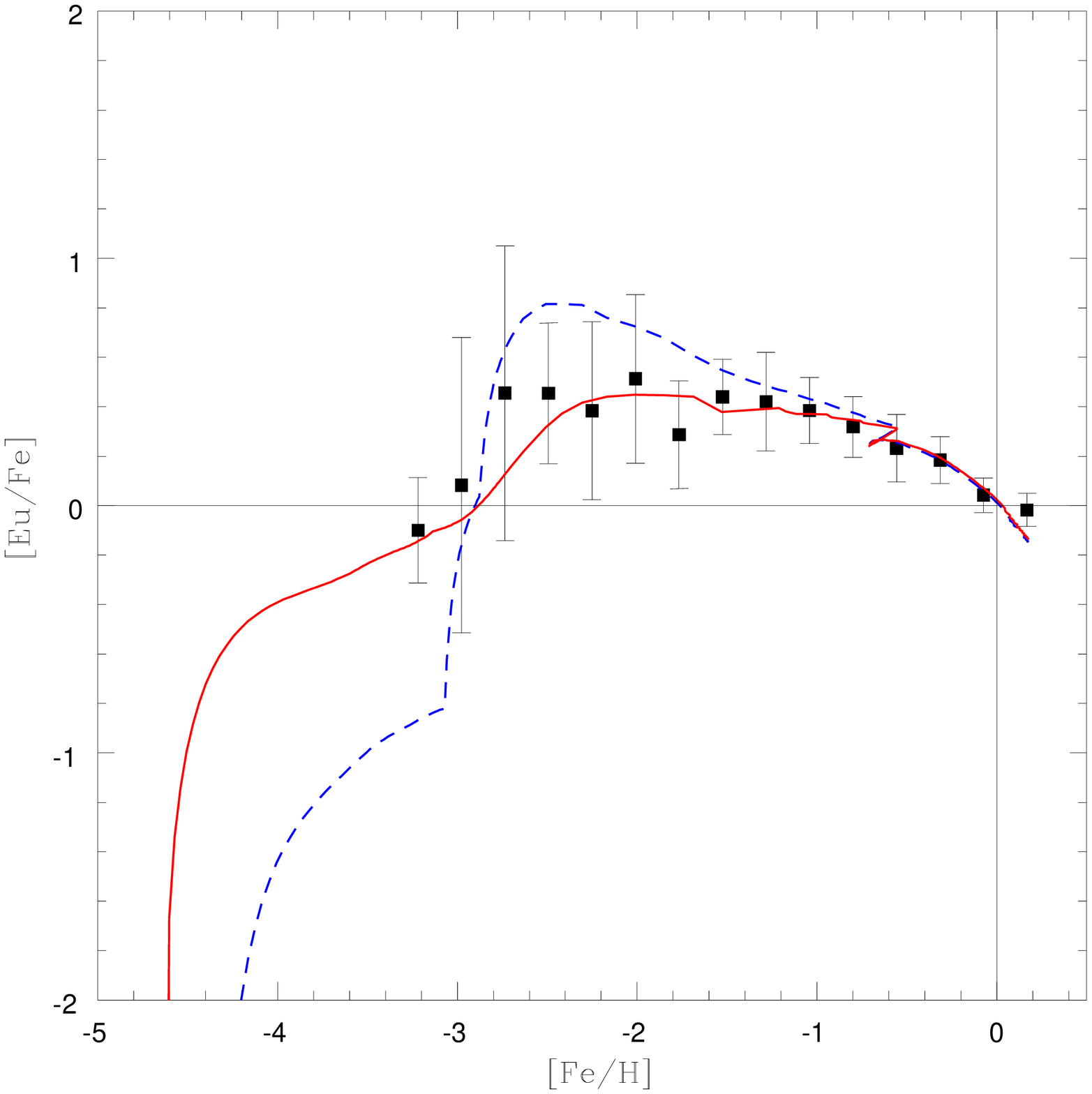}
\caption{The data are as in Fig. \ref{ISHI} and in this Fig.
 we show in solid line the results of the model 1 and in dashed line the 
results of model 2
 (models  are described in table \ref{model}).}\label{best2}
\end{center}
\end{figure}

We have obtained similar results comparing the trend of the abundance
of [Eu/Fe] versus [Fe/H] with the three models of 
Ishimaru et al.(2004) (model 4, 5 and 6 in table \ref{model}).
The chemical evolution of this only r-process element
is shown in Fig. \ref{ISHI} and we want to 
underline again that they used a different chemical model.
Again neither model 4 explains the low metallicity abundances
nor  model 5 and 6 well fit the trend of the data.
In Fig. \ref{best2} we show the results of the models 1 and
2 in this case for [Eu/Fe] versus [Fe/H]. The trend of the data
is well followed by both models from low metallicity
to the solar metallicity.

In table  \ref{bariumsol}  we show the predicted 
solar abundances 
of Eu and Ba for all our  models compared to the solar abundances
 by  Grevesse \& Sauval (1998). We have put in the table
also the predicted s-process fraction in the Barium solar abundance.
The results of almost all our models are in good agreement with the 
solar abundances with the exception of model 5 which underpredicts 
the Eu abundance by a factor of nearly 2.
Note that we predict a slightly different s-process fraction 
(nearly 60\% instead of 80\%) compared with the s-process fraction
obtained by previous authors (Travaglio et al. 1999, Arlandini et al. 1999,
 Raiteri et al. 1992 and  K\" appeler et al. 1989).
This different result is again due to the adopted chemical evolution model, 
as discussed previously.

\begin{figure}
\begin{center}
\includegraphics[width=0.5\textwidth]{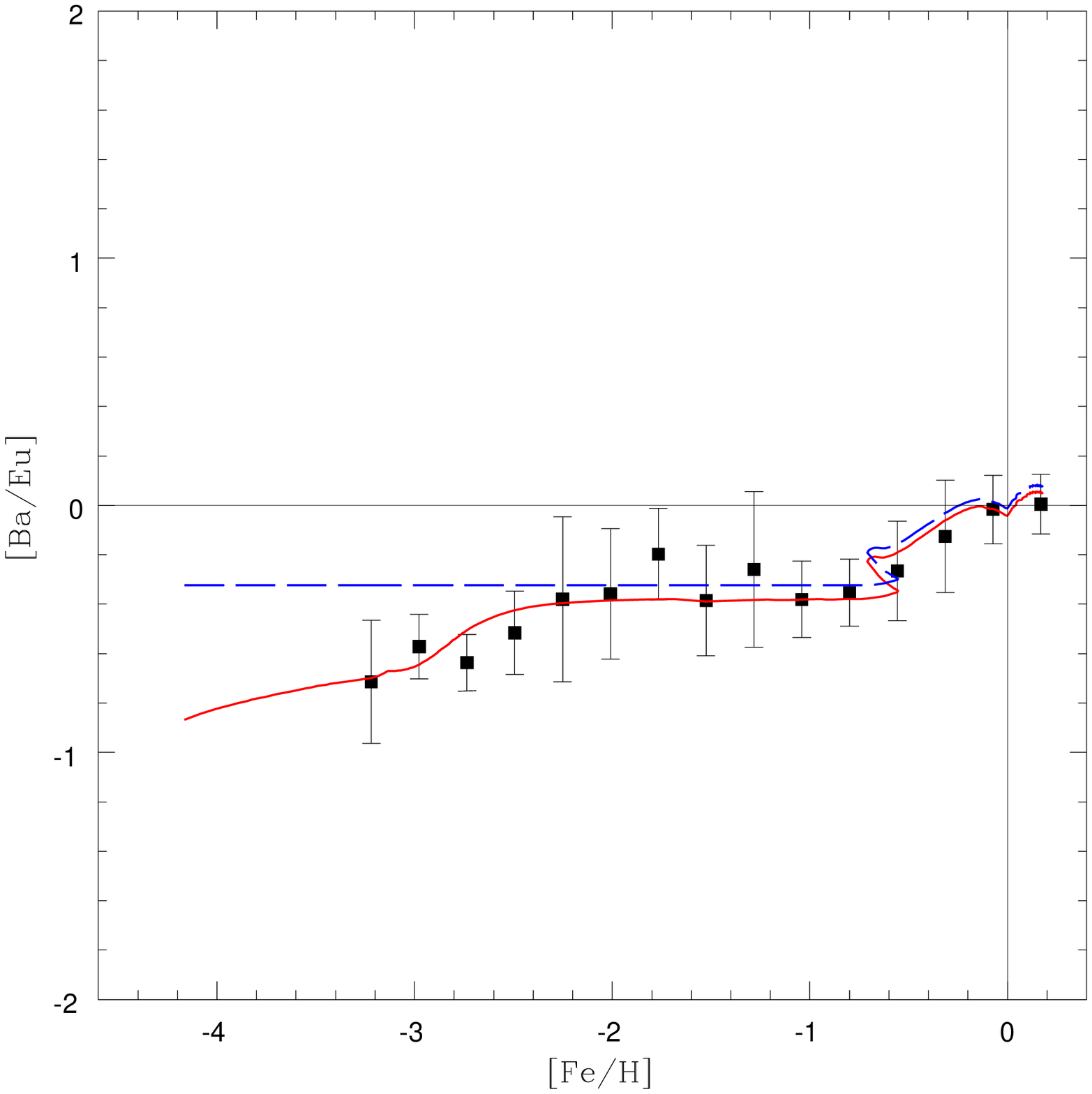}
\caption{In this Fig. we show the ratio of [Ba/Eu] versus [Fe/H]. The squares are
  the mean values of the data bins described 
in the table \ref{meanEu}. As error bars we consider the standard deviation (see table \ref{meanEu}).
 The results of model 1 are rappresented in solid line, the results of  model 2 in long dashed line
 (models  are described in table \ref{model}).}\label{BaEu}
\end{center}
\end{figure}

\begin{figure}
\begin{center}
\includegraphics[width=0.5\textwidth]{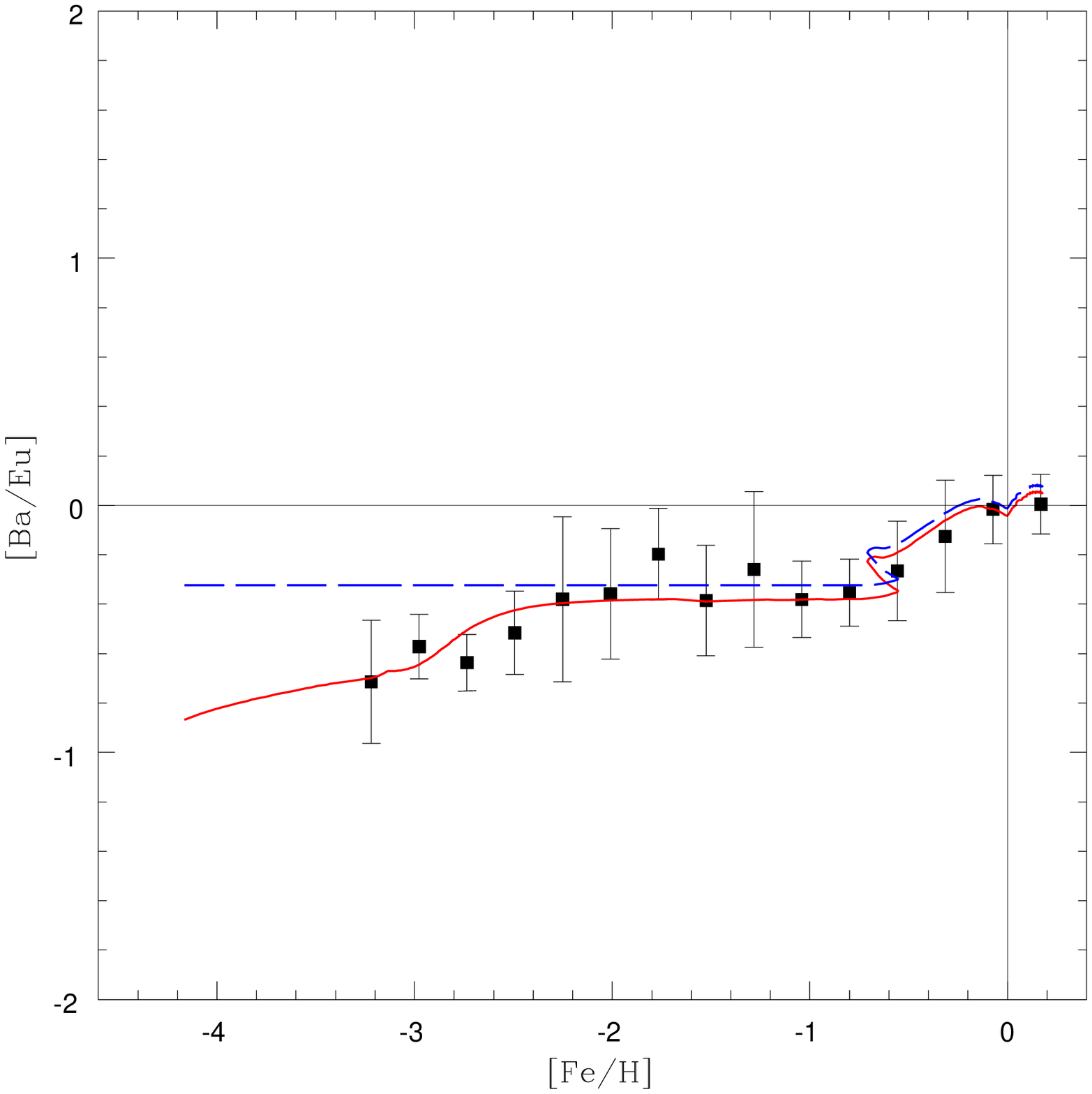}
\caption{In this Fig. we show the ratio [Ba/Eu] versus [Ba/H]. The squares are
  the mean values of the data bins described 
in the table \ref{meanBaH}. As error bars we consider the standard deviation (see table \ref{meanBaH}).
The results of model 1 are rappresente in solid line, the results of model 2 in
 long dashed line (models  are described in table \ref{model}). }\label{BaEu2}
\end{center}
\end{figure}

If we look at Fig. \ref{BaEu} where we have plotted
the abundances of [Ba/Eu] versus [Fe/H] and at 
Fig. \ref{BaEu2}, where is plotted [Ba/Eu] versus [Ba/H],
we note three important features.
The first is that the spread, that we can infer in these 
plots from the standard deviation of each bin, is smaller 
if we use the [Ba/H] ratio in the x axis; 
the second feature is that it is evident from the data
 that there is a plateau in the [Ba/Eu] ratio
before the production of s-process Ba by the low intermediate
mass stars starts to be non negligible at $[Fe/H]\sim-1$ and $[Ba/H]\sim-0.8$;
 finally, the timescale of rise of the [Ba/Eu] value, due to the production of Ba by low intermediate mass stars,
 is very well reproduced by our model.

The value of [Ba/Eu] at low metallicity is an important sign to understand
the fraction of slow processed Ba in the solar abundance
In fact, if the ratio  $\frac{Ba_{rapid}}{Eu}$
 has a constant value over all the cosmic time,
 then it must have the same
value also at the solar system formation time
(if we do not want to add some peculiar effect during
the last part of the Galaxy evolution).

In this case the Ba s-process fraction is simply:

\begin{displaymath}
\frac{Ba_{slow}}{Ba_{total}}=1-10^{[\frac{Ba_{rapid}}{Eu}]}
\end{displaymath}

Since we have a mean value for [Ba/Eu] versus [Fe/H] in the
range $-3<[Fe/H]<-1$ of -0.44 (taken from the mean
value in the bins which fall in that range)
and a similar value for [Ba/Eu] versus [Ba/H] in the range
 $-4<[Ba/H]<-0.8$ of -0.41 (computed in the same way as above), 
then it turns out that  the s-process fraction for Barium
slow processed has to be less than the claimed 80\% with
a value of $\sim 60\%$.

Finally, we want to say that the spread in the ratio of
[Ba/Eu] both versus [Fe/H] and [Ba/H] is 
lower than the spread
in [Ba/Fe] and [Eu/Fe], in particular when using as evolutionary tracer the [Ba/H].

In fact, considering the computed standard deviations  as spread tracers,
 where the spread for [Ba/Fe] and [Eu/Fe] is higher ($[Fe/H]\sim -3$),
their standard deviations are greater than 0.6 dex whereas  the standard deviations
 for [Ba/Eu] is less than 0.15 dex.

For this reason we believe that the mechanism which
produces the observational spread does not affect
the ratio of these two elements.
In our picture the explanation of the smaller spread in the ratio
 of [Ba/Eu] is that the site of production of these
two elements is the same, the neutronized shell close to the
mass cut in a SNII (see Woosley et al. 1994). What changes
could be the amount of the neutronized material which each
massive star expells during the SNII explosion. In fact, 
the mass cut and also the ejected neutronized material,
are still uncertain quantities and usually they are considered as parameters
in the nucleosynthesis codes for massive stars (see Rauscher et al. 2002, 
Woosley \& Weaver 1995, Woosley et al. 1994).

\subsection{Comparison with inhomogeneus chemical evolution models}\label{inomog}

Our chemical evolution model is a model where instantaneous mixing 
is assumed, (ie. shorter than the 
timestep of integration of the equations), therefore it is only able to 
follow the general trends of the abundance ratios as a function 
of [Fe/H] found in the metal-poor stars.
Inhomogeneous models assume that supernovae are able to pollute a 
small region of  gas 
surrounding them, and depending on the mass of the supernova 
(hence the yields) and the size of the polluted zone, 
an abundance spread is predicted. An important point is that the 
abundance ratios do not change with the amount 
of gas which is polluted by the supernova. It is therefore interesting 
to look at the range of the abundance ratios 
log(Ba/Fe) that we obtain with our model and compare them with the spread 
found in the observations.

\begin{figure}
\begin{center}
\includegraphics[width=0.5\textwidth]{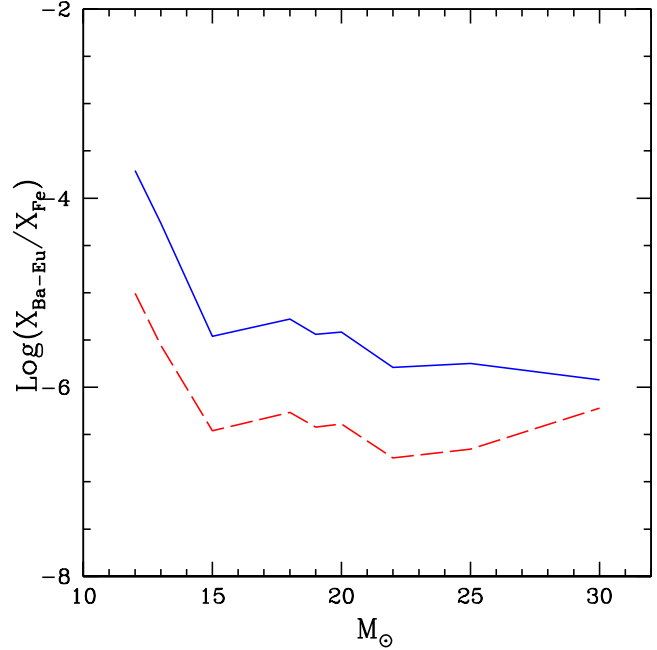}
\caption{ In this Fig  we show the ratio  between the newly produced Ba 
and Fe as a function of the stellar mass for the model 1 yields (solid line); the same for Eu
in long dashed line.}\label{mo}
\end{center}
\end{figure}

In  figure \ref{mo} we show the ratio between the r-process 
elements production and iron production
as a function of stellar mass in the range $12 - 30M_{\odot}$.
This means that the maximum spread which could be found with any 
inhomogeneous model of chemical evolution 
is shown by the range of abundance ratios displayed in 
figure \ref{mo} i.e. about 2 dex.

The observational data instead are  varying in a range which is
of the order of 3 dex  
(see the data in figures \ref{Mm1Ba} for Ba and \ref{Mm1Eu}
for Eu). Therefore, even using an inhomogeneous model with our best 
model yields it would not explain the spread of the data.
 However, by adjusting the production of Ba and Eu only in
 a very narrow interval near 10  $M_{\odot}$, one can account for the observed spread.
This adjustment may even be justified because SNe of 10 $M_{\odot}$ result in O/Mg/Ne 
core collapse and make very little Fe, but r-process elements could be 
produced in the ejecta from the neutron star.
A similar approach  has been used by the inhomogeneous model
 of Argast et al. (2004) and by Ishimaru et al. (2004).

From the results obtained in this work, it is not possible to reproduce, 
even with an inhomogeneous chemical history, such a large spread in the data. 
It is possible that this discrepancy  might be due to the existence
of another  parameter besides the initial mass determining the SN II yields,
such as perhaps the initial angular momentum. 
In the following section, we explore, under the assumption that
the same massive stars contribute both to Fe and r-process synthesis, 
i.e. without any decoupling, which yield ratios a supernova should  
produce to explain both the very high and the very low [r-process/Fe] 
found in the sample of observed halo stars.

\subsection{Upper and lower limit to the r-process production}\label{limits}

\begin{table*}

\caption{Solar abundances of Ba and Eu, as predicted by our models, 
compared with 
the observed ones from  Grevesse \& Sauval (1998).}\label{bariumsol}

\begin{tabular}{|c|c|c|c|c|c|}
\hline\hline

Mod               & $(X_{Ba})_{pr}$        & \%$Ba_{s}/Ba$ & $X_{Ba_{\odot}}$  &$(X_{Eu})_{pr}$       & $X_{Eu_{\odot}}$      \\        
\hline\hline                                                                                                                            
1                 & $1.55\cdot 10^{-8}$    &54\%           &$1.62\cdot 10^{-8}$&$4.06\cdot 10^{-10}$  & $3.84\cdot 10^{-10}$  \\         
\hline                                                                                                                          
2                 & $1.62\cdot 10^{-8}$    &51\%           &                   & $3.96\cdot 10^{-10}$ &                       \\         
\hline                                                                                                                          
3                 & $1.64\cdot 10^{-8}$    &44\%           &                   & As model 1           &                       \\        
\hline                                                                                                                          
4                 & As model 1             &As model 1     &                   & $4.48.\cdot 10^{-10}$&                       \\        
\hline                                                                                                                          
5                 & As model 1             &As model 1     &                   & $1.86\cdot 10^{-10}$ &                       \\        
\hline                                                                                                                          
6                 & As model 1             &As model 1     &                   & $2.84\cdot 10^{-10}$ &                       \\        
\hline\hline

\end{tabular}

\end{table*}

The purpose of this Sect. is to give upper and lower limits
to the yields in order to reproduce the observed spread at low metallicities 
for Ba and Eu.
We are well aware that an inhomogeneous model would provide better 
predictions about the dispersion in the [r-process/Fe] ratios if due to 
yield variations, but we also believe that it is still useful to study the 
effect of the yield variations by means of our model.

First we explore the range of variations of the yields as functions
of the  stellar mass. To do this we have worked on the model 1:  in particular,
we  have modified
the yields of the model 1 for both elements (Ba and Eu), leaving untouched 
the s-process yields and changing only the yields of the r-process.

1Max and 1min and their characteristics are summarized in the table \ref{Mm}.

\begin{table*}

\caption{The stellar yields for model 1Max and 1Min for Barium and Europium in massive stars (r-process)
in the case of a primary origin.} \label{Mm}

\centering
\begin{minipage}{90mm}

\begin{tabular}{|c|c|c|c|c|}
\hline

             & Model 1Max         &    & Model 1Min         &    \\     
\hline\hline
$M_{star}$   & $ X_{Ba}^{new}$    & Factor & $ X_{Ba}^{new}$     & Factor\\

12.          & 1.35$\cdot10^{-6}$ & 1.5    & 4.50$\cdot10^{-7}$  & 0.5\\ 
$ < 15$.     & 4.50$\cdot10^{-8}$ & 1.5    & 1.50$\cdot10^{-8}$  & 0.5\\   
$ \ge 15$    & 3.00$\cdot10^{-7}$ & 10.    & 1.50$\cdot10^{-9}$  & 0.05\\
30.          & 1.00$\cdot10^{-8}$ & 10.    & 5.00$\cdot10^{-11}$ & 0.05\\ 

\hline\hline         
                                            
$M_{star}$   & $ X_{Eu}^{new}$    & Factor& $ X_{Eu}^{new}$     & Factor\\
12.          & 4.50$\cdot10^{-8}$        & 1.    & 2.25$\cdot10^{-8}$         & 0.5 \\ 
$ < 15$.     & 3.00$\cdot10^{-9}$        & 1.    & 1.50$\cdot10^{-9}$         & 0.5 \\   
$ \ge 15$    & 3.00$\cdot10^{-8}$        & 10.   & 1.50$\cdot10^{-10}$        & 0.05\\
30.          & 5.00$\cdot10^{-9}$        & 10.   & 2.50$\cdot10^{-11}$ & 0.05\\

\hline\hline

\end{tabular}

\end{minipage}

\end{table*}

\begin{figure}
\begin{center}
\includegraphics[width=0.5\textwidth]{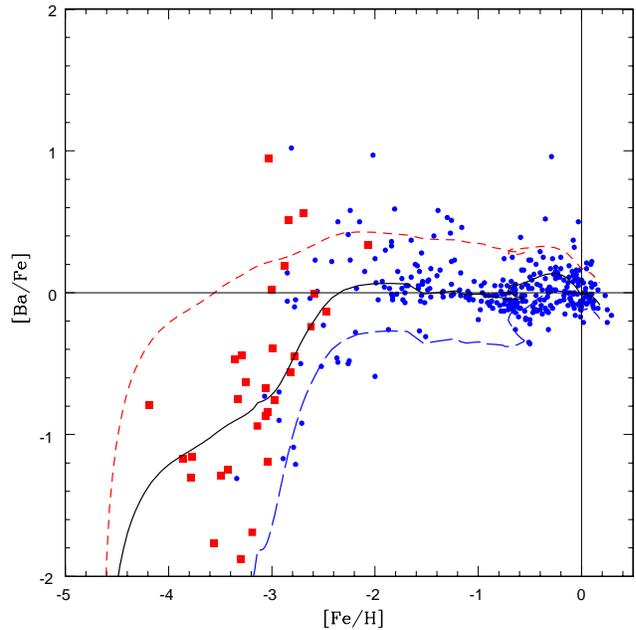}
\caption{In this Fig. we show the ratio [Ba/Fe] versus [Fe/H] for  the data by 
Fran\c cois et al. (2005) (filled squares)  and
 for the other observational data (see Sect. 2 in the text, the filled circles).
The solid line is the prediction of model 1, the short dashed line the prediction of model 1Max
and  the long dashed line the prediction of model 1min.}
\label{Mm1Ba}
\end{center}
\end{figure}

\begin{figure}
\begin{center}
\includegraphics[width=0.5\textwidth]{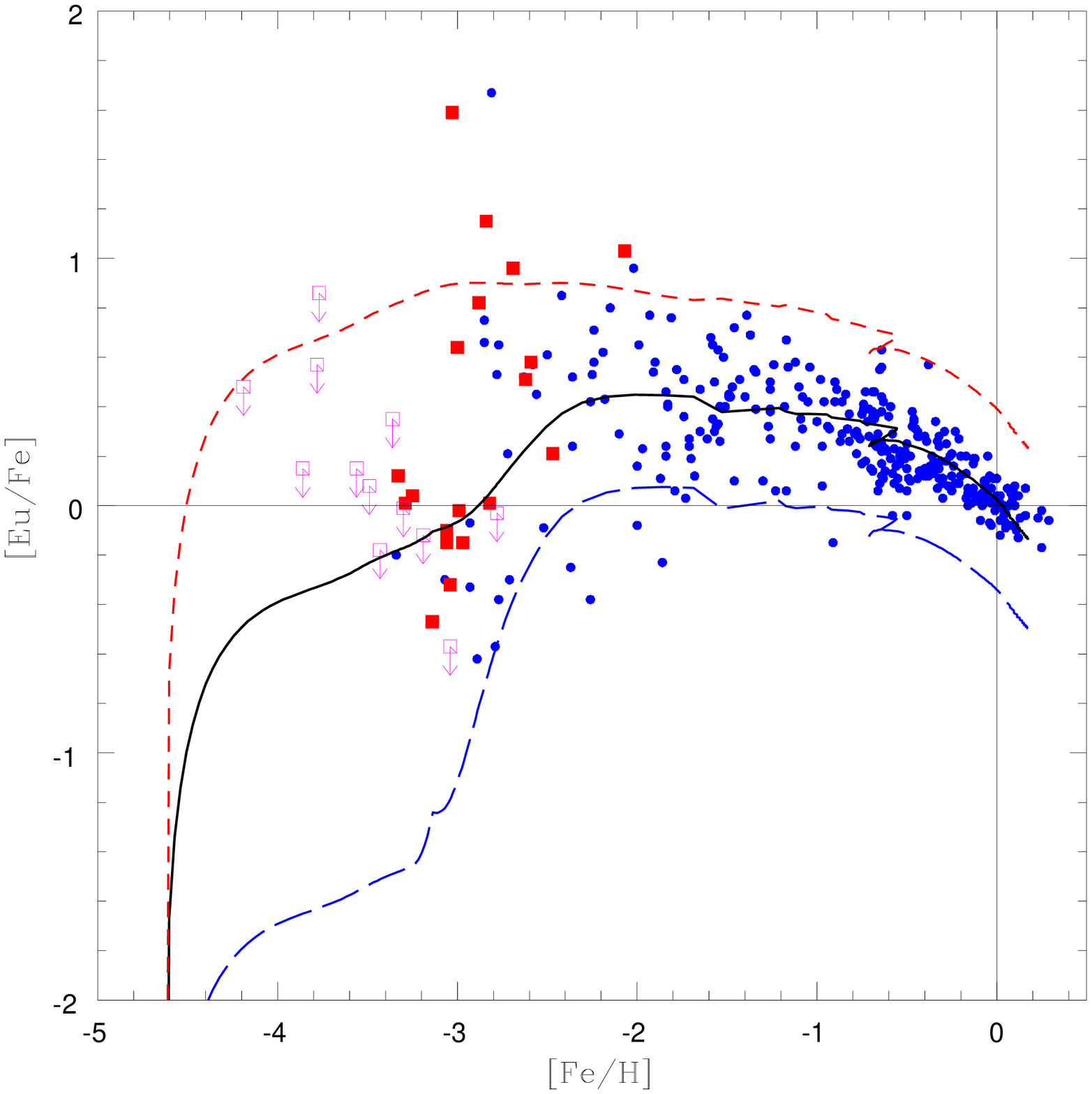}
\caption{In this Fig. we show the ratio [Eu/Fe] versus the ratio [Fe/H]. The data by 
Fran\c cois et al. (2005) are with filled squares, wheras the open squares
are only upper limits again from the work of Fran\c cois et al. (2005). The filled circles are data 
by other observational works (see Sect. 2 in the text).
The solid line is the prediction of model 1, the short dashed line the prediction of model 1Max
and  the long dashed line the prediction of model 1min.}
\label{Mm1Eu}
\end{center}
\end{figure}

\begin{figure}
\begin{center}
\includegraphics[width=0.5\textwidth]{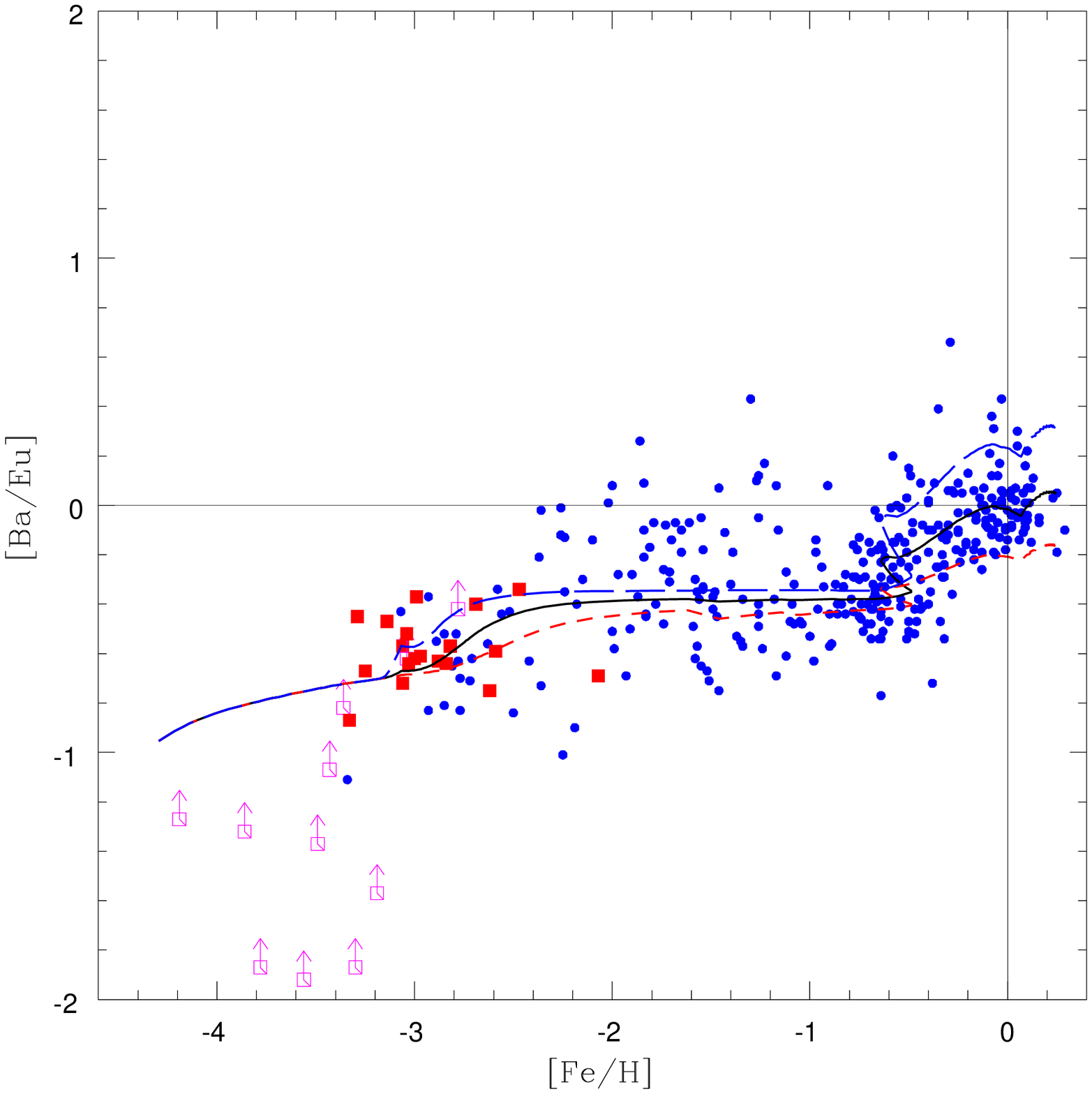}
\caption{In this Fig. we show the ratio [Ba/Eu] versus the ratio [Fe/H]. The data by 
Fran\c cois et al. (2005) are filled squares, wheras the open squares
are only lower limits again from the work of Fran\c cois et al. (2005). The filled circles are data 
by other observational works (see Sect. 2 in the text).
The solid line is the prediction of model 1, the short dashed line the prediction of model 1Max
and  the long dashed line the prediction of model 1min.}
\label{Mm1BaEu}
\end{center}
\end{figure}

\begin{figure}
\begin{center}
\includegraphics[width=0.5\textwidth]{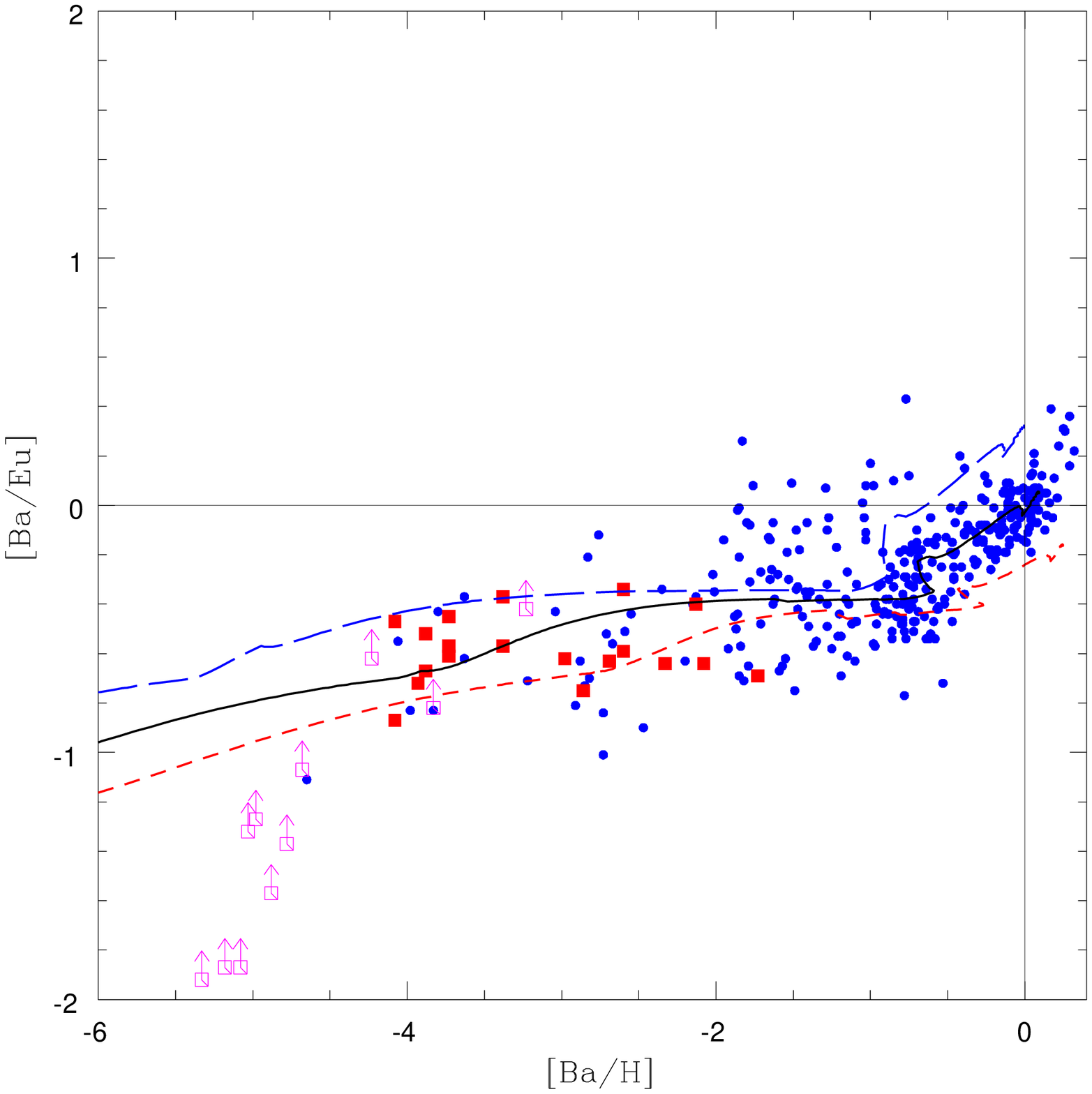}
\caption{As in Fig. \ref{Mm1BaEu} but  for the ratio [Ba/Eu] versus [Ba/H].}
\label{Mm1BaEu2}
\end{center}
\end{figure}

In Fig. \ref{Mm1Ba} and \ref{Mm1Eu}  we plot ratios [Ba/Fe] vs [Fe/H] and [Eu/Fe] vs [Fe/H]
 respectively, for the new models 1Max 1min and 1 compared to the observational data;
 we show the same plot  for the ratios [Ba/Eu] vs [Fe/H] in  Fig.\ref{Mm1BaEu} and
and for [Ba/Eu] versus [Ba/H]  in Fig. \ref{Mm1BaEu2}.

We can deduce from  these upper and lower limit models that the large observed spread 
could be also due to a  different production of heavy elements among massive stars ($>15M_{\odot}$).
This type of stars could produce different amounts of these elements
independently of the mass.
As we have introduced in the previous subsect. (\ref{trends}), it is possible to link this fact 
with the problems of mass cut and the fall back during the explosion of
a SNII. If these elements are produced in a shell close to the iron core of the star, differences in the 
explosion behaviour can give rise to a different quantity of r-process elements  expelled by the star.

In this way we are able to explain the presence of the spread for the
heavy elements and the absence of the same spread for example in the $\alpha$-elements.
 In fact, the $\alpha$-elements are produced mostly
during the hydrostatic burning of massive stars and then  ejected by the explosion.

Another approach can be followed to derive
again  upper and lower limit for the model 
by changing the yields as functions of metallicity.
The model 2, which is the model with yields independent of the mass and depending only on metallicity, will be our test model. 
In particular, Model 2 assumes for the massive
stars different yields for Ba and Eu in three ranges of  metallicity
(see table \ref{model}).

The new prescriptions  for both Ba and Eu are summarized in table \ref{rBa3}.

\begin{table*}

\caption{The stellar  yields of model 2Max and model 2min for Ba and Eu 
in massive stars (r-process).} \label{rBa3}

\centering

\begin{minipage}{90mm}

\begin{tabular}{|c|c|c|c|c|}

\hline\hline

 $Z_{star}$                &  $Z_{star}$                     & $X_{Ba}^{new}$     & $ X_{Eu}^{new}$ \\
 model 2Max                &  model 2min                     & $10-25M_{\odot}$ & $10-25M_{\odot}$ \\
                                                         
\hline                           
           never           &            $Z<8\cdot10^{-6}$.   & 1.00$\cdot10^{-8}$ &  5.00$\cdot10^{-10}$  \\ 
      $Z<1\cdot10^{-5}$    & $8\cdot10^{-6}<Z<1\cdot10^{-5}$ & 1.00$\cdot10^{-6}$ &  5.00$\cdot10^{-8}$  \\   
      $Z>1\cdot10^{-5}$    &            $Z>1\cdot10^{-5}$    & 1.60$\cdot10^{-7}$ &  8.00$\cdot10^{-9}$  \\ 

\hline\hline

\end{tabular}

\end{minipage}

\end{table*}

\begin{figure}
\begin{center}
\includegraphics[width=0.5\textwidth]{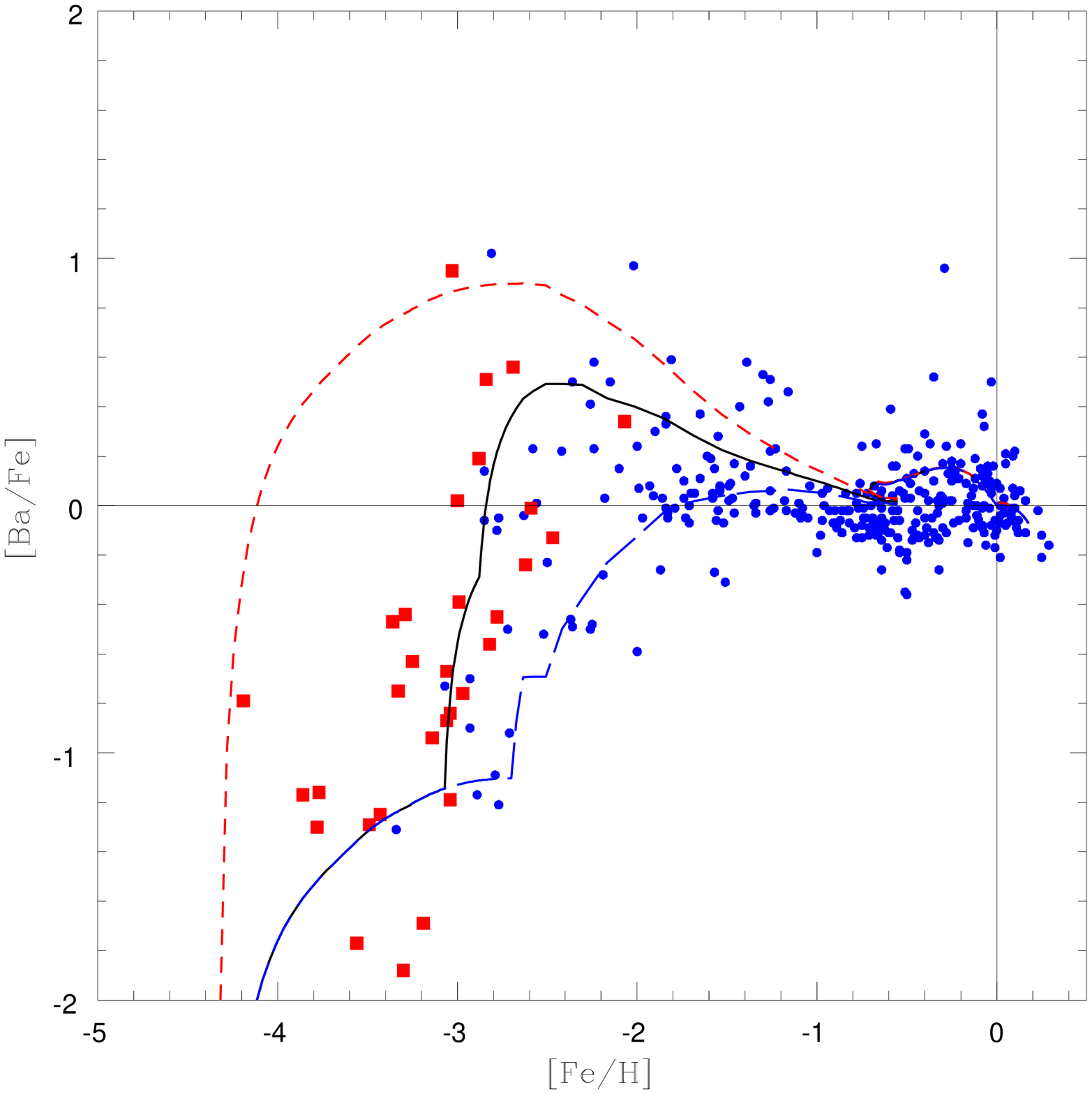}
\caption{In this Fig. we show the ratio of [Ba/Fe] versus [Fe/H]. The data are as in Fig. \ref{Mm1Ba}.
The solid line is the prediction of model 2, the short dashed line the prediction of model 2Max
and  the long dashed line the prediction of model 2min.}\label{Mm2Ba}
\end{center}
\end{figure}

\begin{figure}
\begin{center}
\includegraphics[width=0.5\textwidth]{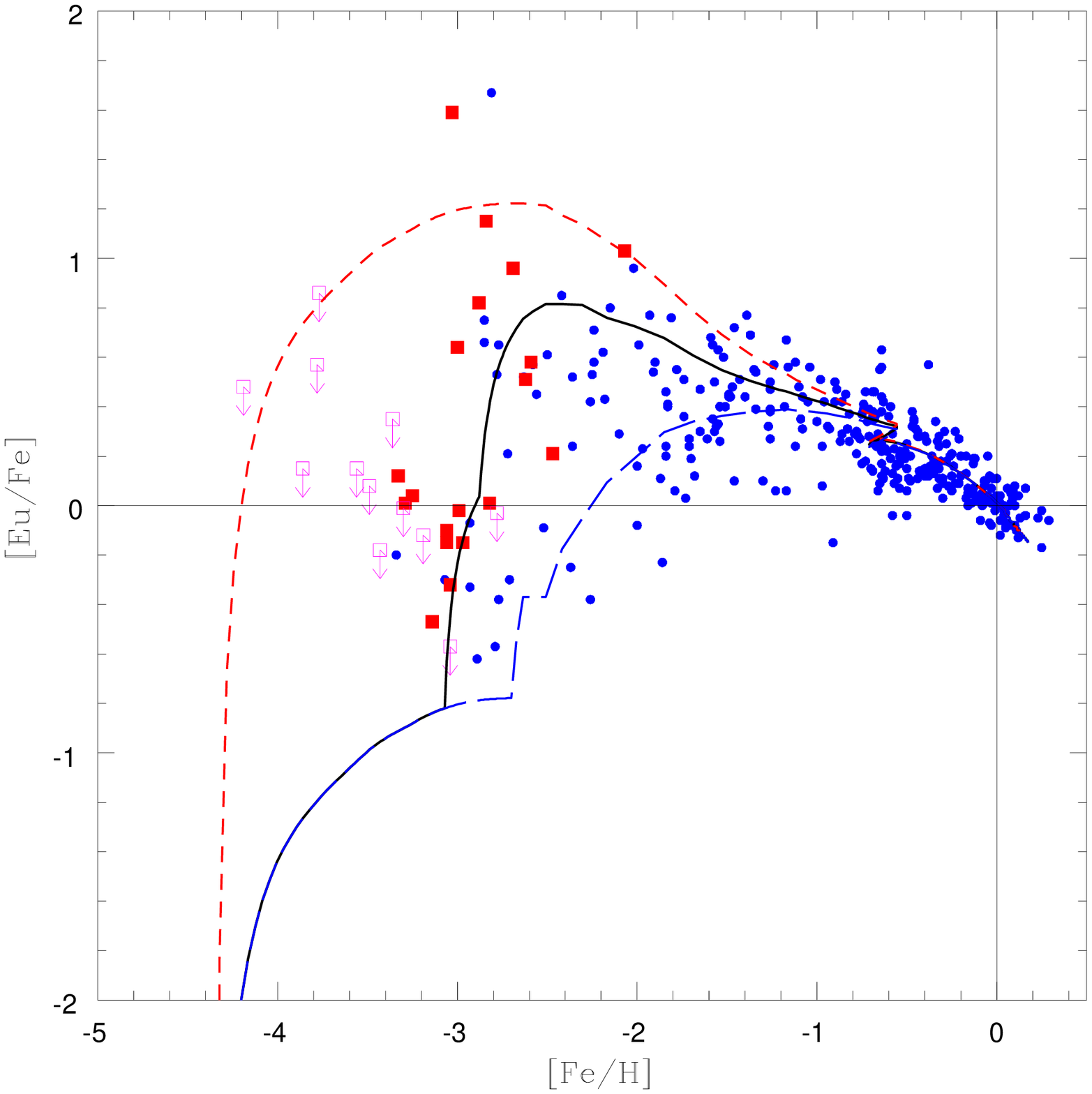}
\caption{In this Fig. we show the ratio [Eu/Fe] versus the ratio [Fe/H]. The data are as in Fig. \ref{Mm1Eu}.
The solid line is the prediction of model 2, the short dashed line the prediction of model 2Max
and  the long dashed line the prediction of model 2min.}\label{Mm2Eu}
\end{center}
\end{figure}

In Fig. \ref{Mm2Ba} and \ref{Mm2Eu}  are shown the results of these two models (2Max and 2min) 
and of the original model 2 for [Ba/Fe] vs [Fe/H] and  for [Eu/Fe] versus [Fe/H] 
respectively, compared to the observational data.

The results are very interesting. In fact, changing the
central range of metallicity, in which there is an enhancement 
of the production of Ba and Eu, it is possible to produce the upper and the 
lower limits.
These two new models envelope the majority of the data at low metallicities.
At higher metallicity the two models overlap the best model and so they do not
contain all the spread in this part of the plot but most of them could 
be explained inside the typical observational error of 0.1 dex.

\section{Conclusions}

The main goal of this work was to follow the evolution of Ba and Eu by 
means of a chemical evolution model well reproducing the abundance trends 
for other elements.
To do that we used the Chiappini et al. (1997) model in its latest version 
as described in Chiappini et al. (2003).
To reach this goal we have used empirical yields for stars with mass 
$>8M_{\odot}$, producing r-process elements.
In fact for the r-process elements there are not solid theoretical yields, 
since the mechanism which is involved in their production, 
the so called r-process, is still not well understood.
We conclude that Ba needs two components: a s-process main component 
originating in low mass stars plus a r-component originating in stars in 
the range 
10-30$M_{\odot}$. This range is different from the one suggested by Travaglio 
et al. (1999) and
is obtained by requiring the best fit of the new data.
For Eu we estimated that it is mainly produced by an r-process and that stars in 
the same mass range 10-30$M_{\odot}$ should be considered as the progenitors of this 
element.

The nearly constant value of the ratio [Ba/Eu] produced in massive stars
by the r-process, can be used to estimate  the fraction of Barium in the solar abundance
produced  by  the slow process. We have obtained in this
way a fraction that is different from the previous results: 60\% instead of 80\%.

We found that the yields ratios which we obtained to reproduce the trends are not able to explain 
the large spread found by the observations even with the use of an inhomogeneous model. 
This implies or the decoupling of the production r-process elements and Fe or a variation of the
 yields as a function of metallicity for example. It could also that the yields of the r-process 
are not only a function of the mass of the progenitor.

The yields that we derived and even the fact that the ratio of the r- process production  
of Eu and Ba seems to be be nearly constant could be very useful to studies involving nucleosynthesis 
in stellar models and even to nuclear physics studies.

\section{Acknowledgments}
We thank Francesco Calura and Cristina Chiappini 
for several useful comments.
GC and FM also acknowledge funds from MIUR, COFIN 2003, prot. N. 2003028039.


\end{document}